\documentclass{JHEP3}
\usepackage{bbm}
\usepackage[final]{graphicx}
\usepackage{amsmath}
\usepackage{amsfonts,amsbsy}
\usepackage{amssymb}
\usepackage{mciteplus}

\newcommand{\ii}{{e_i}}

\newcommand{\ex}{e_x}
\newcommand{\ey}{e_y}

\newcommand{\ud}{\, \mathrm{d}}

\newcommand{\trace}{\, \mathrm{Tr} \, }
\newcommand{\R}{\mathrm{Re}}
\renewcommand{\nc}{{N_\mathrm{c}}}

\newcommand{\hc}{\mathrm{\ h.c.\ }}
\newcommand{\nosum}[1]{\textrm{ (no sum over } #1 )}
\newcommand{\lqcd}{\Lambda_{\mathrm{QCD}}}
\newcommand{\as}{\alpha_{\mathrm{s}}}
\newcommand{\Tr}{\mathrm{Tr}}

\newcommand{\nr}[1]{(\ref{#1})} 

\newcommand{\ra}{R_A}
\newcommand{\gev}{\ \textrm{GeV}}
\newcommand{\mev}{\ \textrm{MeV}}

\newcommand{\qs}{Q_\mathrm{s}}

\newcommand{\fig}{fig.~}

\newcommand{\eq}{eq.~}
\newcommand{\se}{sec.~}
\newcommand{\eqs}{eqs.~}

\def\y{{\boldsymbol y}}
\def\x{{\boldsymbol x}}
\def\p{{\boldsymbol p}}
\def\q{{\boldsymbol q}}

\def\x{{\boldsymbol x}}
\def\y{{\boldsymbol y}}

\newcommand{\xt}{\mathbf{x}_T}
\newcommand{\yt}{\mathbf{y}_T}

\newcommand{\pt}{{\mathbf{p}_T}}
\newcommand{\qt}{\mathbf{q}_T}
\newcommand{\kt}{\mathbf{k}_T}

\newcommand{\nabt}{\boldsymbol{\nabla}_T}

\title{Non-perturbative computation of double inclusive gluon 
production in the Glasma}

\author{T. Lappi\\
Physics Dept., P.O. Box 35,
 40014, University of Jyv\"{a}skyl\"{a}  and \\
Helsinki Institute of Physics, P.O. Box 64, 00014 University of Helsinki, Finland
\\ Email: \email{tuomas.lappi@jyu.fi}
}
\author{S. Srednyak\\
Physics Dept., SUNY Stony Brook, Stony Brook, NY 11790 and\\ 
Physics Dept., Bldg. 510A, Brookhaven National Laboratory,
 Upton, NY 11973, USA
\\ Email: \email{stas\_srednyak@yahoo.com}
}
\author{R. Venugopalan \\
Physics Dept., Bldg. 510A, Brookhaven National Laboratory, 
 Upton, NY 11973, USA
\\ Email: \email{rajuv@mac.com}
}

\abstract{The near-side ridge observed in A+A collisions at RHIC has been 
described as arising from the radial flow of Glasma flux tubes formed at 
very early times in the collisions. We investigate the viability of this 
scenario by performing a non-perturbative numerical computation of double inclusive 
gluon production in the Glasma. Our results support the conjecture that 
the range of transverse color screening of correlations determining the
 size of the flux tubes is a semi-hard scale, albeit  with non-trivial
 structure. We discuss our results in the context  of ridge correlations
 in the RHIC heavy ion experiments.}

\begin{document}

\section{Introduction}

At very high energies, the collision of two nuclei can be conveniently described as the collision of two Color Glass Condensates (CGCs)~\cite{McLerran:1994ni,*McLerran:1994ka,*McLerran:1994vd,%
Jalilian-Marian:1997xn,*Jalilian-Marian:1997jx,*Jalilian-Marian:1997gr,%
*Jalilian-Marian:1997dw,*JalilianMarian:1998cb,*Weigert:2000gi,*Iancu:2000hn,%
*Iancu:2001md,*Ferreiro:2001qy,*Iancu:2001ad,*Mueller:2001uk,Iancu:2003xm,*Weigert:2005us}. 
In this description, the wavefunctions of the nuclei are comprised of high occupation number classical gluon fields at small $x$ coupled to static light cone color sources at large $x$. The separation between fields and sources evolves with energy; one obtains evolution equations for multi-parton correlations in the nuclear wavefunctions called JIMWLK renormalization group 
equations~\cite{Jalilian-Marian:1997xn}. After the collision, the coherence of the
 nuclear wavefunction is lost; large $x$ sources are no longer static and become sources for multi-particle production in the forward light 
cone~\cite{Gelis:2006yv,*Gelis:2006cr}.
 For early times $\tau \sim 1/\qs$, where $\qs$ is the saturation scale, 
the non-equilibrium produced matter with  high occupation number of order 
$1/\as$ has remarkable properties~\cite{Kovner:1995ts,Kharzeev:2001ev} and is
 termed the Glasma~\cite{Lappi:2006fp}. For reviews, 
see Ref.~\cite{Gelis:2006dv,*Gelis:2007kn}.

Recently, high energy factorization theorems were derived for 
inclusive multi-gluon  production in a rapidity interval 
$\Delta y \lesssim 1/\as$~\cite{Gelis:2008rw,Gelis:2008ad} 
in A+A collisions. The result can be expressed very compactly 
as~\cite{Gelis:2008ad}
\begin{equation}
\left<\frac{\ud^n  N_n}{\ud^3\p_1\cdots \ud^3\p_n}\right>_{_{\rm LLog}}
 =
  \int \big[D\rho_1\big]\big[D\rho_2\big]\;
  Z_{y}\big[\rho_1\big]\,
  Z_{y}\big[\rho_2\big]\;
  \left.\frac{\ud N}{\ud^3\p_1}\right|_{_{\rm LO}}\cdots
  \left.\frac{\ud N}{\ud^3\p_n}\right|_{_{\rm LO}}\; .
\label{eq:ngluon-LLog}
\end{equation}
The $Z$'s are gauge invariant weight functionals that describe the 
distribution of color sources at the rapidity of interest. They are 
obtained in full generality by evolving the 
JIMWLK equations from an initial rapidity close to the 
beam rapidity. In the large $\nc$ limit, the weight functionals $Z$ 
can instead be obtained from the simpler mean field Balitsky--Kovchegov 
(BK)~\cite{Balitsky:1995ub,*Balitsky:1998kc,*Balitsky:1998ya,*Kovchegov:1999yj,*Kovchegov:1999ua,*Balitsky:2001re} 
equation. In this case, they can be represented as non-local Gaussian distributions in 
the sources~\cite{Blaizot:2004wv}. For large nuclei, without significant small $x$ 
evolution, one recovers the local Gaussian distributions of the McLerran-Venugopalan 
(MV) model~\cite{McLerran:1994ni,*McLerran:1994ka,*McLerran:1994vd}. 
We emphasize that the validity of \eq\nr{eq:ngluon-LLog} is restricted to the kinematics where all the produced particles 
are measured within a rapidity interval $ \lesssim 1/\as$ from each other, so that we can evaluate
$Z$ at this same rapidity $y_1 \approx \cdots \approx y_n \approx y$. The generalization to larger rapidity separations is non-trivial 
because one needs to account for the gluon radiation in the region between the tagged gluons. A formalism describing arbitrarily long range rapidity 
separations has been developed recently~\cite{Gelis:2008sz,*Lappi:2009fq}.
 For simplicity, we shall not consider it further here. 

The leading order single particle distributions in 
\eq(\ref{eq:ngluon-LLog}) are given by 
\begin{multline}
\left.E_\p{\frac{\ud N}{\ud^3\p}}\right|_{_{\rm LO}} \Big[\rho_1,\rho_2\Big]=\frac{1}{16\pi^3}
\lim_{x_0\to+\infty}\int \ud^3\x \, \ud^3\y
\;e^{ip\cdot(x-y)}
\;(\partial_x^0-iE_\p)(\partial_y^0+iE_\p)
\\
\times\sum_{\lambda}
\epsilon_\lambda^\mu(\p)\epsilon_\lambda^\nu(\p)\;
A_\mu^{a, \rm cl.}[\rho_1,\rho_2](x)\;A_\nu^{a, \rm cl.}[\rho_1,\rho_2](y)\; .
\label{eq:AA}
\end{multline}
For each configuration of sources $\rho_1$ and $\rho_2$ respectively, of each of the nuclei, 
one can solve classical Yang--Mills equations to compute the gauge fields 
$A_\mu^{\rm cl.}[\rho_1,\rho_2]$ in the forward light cone~\cite{Krasnitz:1998ns,*Krasnitz:1999wc,*Krasnitz:2000gz,Krasnitz:2001qu,Krasnitz:2003jw,Lappi:2003bi,*Lappi:2006hq,Krasnitz:2002ng,*Krasnitz:2002mn,*Lappi:2006xc}. 
When our expression for the corresponding single inclusive distribution is substituted 
in \eq(\ref{eq:ngluon-LLog}), and the distributions are averaged over with the distributions $Z$, one has determined from first principles (to all orders in perturbation theory and to leading logarithmic accuracy\footnote{Note that what we mean by leading log of $x$ here means resumming the leading dependence in rapidity between the  observed gluons and the projectiles; not between the different produced gluons.} 
in $x$), the n-gluon inclusive distribution in high energy A+A collisions at proper times 
$\tau\sim 1/\qs$. As noted previously,  \eq(\ref{eq:ngluon-LLog}) is valid only for 
$\Delta Y\lesssim 1/\as\sim 3-5$ 
units of rapidity in A+A collisions at RHIC and the LHC respectively.

In the nuclei, before the collision, the typical range of color correlations is of the order of the inverse saturation scale $\qs^{-1}$, where 
${\qs}^{-1} < {\lqcd}^{-1}$. The saturation scale at a given transverse position in the nucleus depends on the two dimensional transverse projection of the nuclear matter distribution. In \eq(\ref{eq:ngluon-LLog}), it appears in the initial conditions for the $Z$ functionals; the energy evolution of the saturation scale is determined by the JIMWLK renormalization group equations.   Because the saturation scales in the two nuclei are the only scales in the problem (besides nuclear radii), their properties determine the energy and centrality dependence of the inclusive observables.  

The expression in \eq(\ref{eq:ngluon-LLog}) is remarkable because it suggests that in 
a single event--corresponding to a particular configuration of color sources--the 
leading contribution is from $n$ tagged gluons that are produced independently. 
The coherence in $n$-gluon emission is generated by averaging over color sources 
that vary from event to event. Because the range of color correlations in the 
transverse plane is of order $1/\qs$, 
the high energy factorization formalism suggests an intuitive picture 
of correlated multiparticle production arising from event by event 
fluctuations in particle production from $\sim \ra^2 / 1/\qs^2 = \ra^2 \qs^2$ 
color flux tubes of size $1/\qs$. 

The color flux tubes, called Glasma flux tubes, are approximately boost 
invariant in rapidity because of the underlying boost invariance of the 
classical fields\footnote{For rapidity separations $\gtrsim 1/\as$,
 one expects significant violations of boost invariance from quantum
 corrections~\cite{Gelis:2008sz}.}. Besides providing the underlying 
geometrical structure
 for long range rapidity correlations, the Glasma flux tubes also carry 
topological charge~\cite{Kharzeev:2001ev}, which may result in observable 
metastable CP-violating domains~\cite{Kharzeev:2007jp}. While 
\eqs\nr{eq:ngluon-LLog} and~\nr{eq:AA} describe particle
 production from the Glasma flux tubes, they do not describe
 the subsequent final state interactions of the 
produced gluons which become important for times $\tau \gg 1/\qs$. 

If, as widely believed, the produced matter thermalizes by final state 
interactions, these will not significantly alter long range rapidity 
correlations because  the effects of fragmentation, hadronization or
resonance decays are typically restricted to $\Delta y \approx 1 \ll 1/\as$.
However, radial flow will have a significant 
effect on the observed angular correlations. This is because particles 
produced isotropically in a given flux tube will be correlated by the 
radial outward hydrodynamic flow of the flux tubes. Ideas on the angular
 collimation of particle distributions by flow were discussed previously
 in the literature~\cite{Voloshin:2003ud,Shuryak:2007fu}. When combined
 with the long range rapidity correlations provided by the Glasma flux 
tubes, they provide a natural 
explanation~\cite{Dumitru:2008wn,Gavin:2008ev,*Moschelli:2009tg}
 of  ``ridge'' like 
structures~\cite{Adams:2005ph,Wang:2004kfa,*Adare:2008cq,*Alver:2008gk} 
 seen in the nearside spectrum of  two particle correlated hadron pairs
 in A+A collisions. A similar structure is seen in 
three hadron correlations as well~\cite{Netrakanti:2008jw,*Abelev:2008nd}.

These structures were first seen in near side events with prominent 
jet like structures,  where  the spectrum of associated particles is 
observed to be collimated in the azimuthal separation $\Delta \varphi$ 
relative to the jet and shows a nearly constant amplitude in the
strength of the pseudo-rapidity correlation $\Delta \eta$ up to 
$\Delta \eta\sim 1.5$~\cite{Adams:2005ph}. The name ``ridge'' follows 
from the visual appearance of these structures as an extended mountain
 ridge in the $\Delta \eta$-$\Delta \varphi$ plane associated with a narrow 
jet peak. This collimated correlation persists up to 
$\Delta \eta ~\sim 4$~\cite{Molnar:2007wy,*Wenger:2008ts}. An important feature 
of ridge correlations is that the above described structure is seen in two particle 
correlations without a jet trigger and persists without significant modification for
 the triggered events~\cite{Adams:2004pa,*Abelev:2009qa}. These events include all
 hadrons with momenta $p_T \geq 150\mev$. In such events, a sharp rise in the 
amplitude of the ridge is seen~\cite{Adams:2004pa,Daugherity:2008su} in going 
from peripheral to central collisions.  A number of theoretical models have been
 put forth to explain these ridge 
correlations~\cite{Shuryak:2007fu,Armesto:2004pt,*Majumder:2006wi,*Pantuev:2007sh,*Wong:2008yh,*Chiu:2008ht}.
 For a recent critical evaluation of these models, we refer the reader to 
Ref.~\cite{Nagle:2009wr}. 

In this paper, our primary purpose is to discuss how the Glasma flux tube 
picture arises  {\it ab initio} from solving \eqs\nr{eq:ngluon-LLog}
and \nr{eq:AA}.  In Ref.~\cite{Dumitru:2008wn}, it was argued that 
the two particle correlation function
\begin{equation}
C_2(\p,\q) \equiv \left<\frac{\ud^2 N_2}{\ud y_p \ud^2 \pt \ud y_q \ud^2 \qt }\right> 
-\left<\frac{\ud N}{\ud y_p \ud^2 \pt}\right>\left<\frac{\ud N}{\ud y_qd^2 \qt}
\right>\,,
\label{eq:C2part}
\end{equation}
in the Glasma flux tube picture took on the simple geometrical form
\begin{equation}
\frac{C_2(\p,\q)}{
\left<\frac{\ud N}{\ud y_p \ud^2 \pt}\right>
\left<\frac{\ud N}{\ud y_q \ud^2 \qt} \right>}
=\kappa_2\,\frac{1}{S_\perp \qs^2}\,, 
\label{eq:C2part2}
\end{equation}
where the right hand side of this relation is simply a non-perturbative constant $\kappa_2$ 
multiplied by the ratio of the transverse area of the flux tube to the transverse overlap 
area of the nuclei. This nice geometrical identity however is a conjecture based on a
 perturbative computation whose regime of validity is the kinematic region 
$p_T,q_T \gg \qs$, where one expects additional contributions, besides those
 arising from strong sources, to contribute significantly. Similar arguments, 
based on perturbative computations were used to generalize the result in 
\eq(\ref{eq:C2part2}) to 3-particle correlations~\cite{Dusling:2009ar} 
and subsequently, even $n$-particle correlations~\cite{Gelis:2009wh}. 
In the latter paper, it was shown that the general structure of $n$-gluon emissions is a 
negative binomial distribution. 

Because the geometrical structure of the correlations is so striking and with manifest 
consequences, it is important to establish that is has a validity beyond perturbation 
theory. The perturbative expressions extended to lower momenta are strongly infrared 
divergent so it is not obvious whether they are regulated in that regime by the confinement 
scale of the order of $\lqcd$ or instead a semi-hard scale of order $\qs^{-1}$ arising from
 the weak coupling albeit non--perturbative dynamics of the Glasma fields. We will tackle 
the problem head-on in this paper and investigate the form of \eq(\ref{eq:C2part}) in this 
paper. We will do so by solving the Yang--Mills equations in \eq(\ref{eq:AA}) in full 
generality numerically on a space-time lattice. Our non-perturbative results from the 
lattice simulations are valid\footnote{We emphasize however that at very large momenta there 
will be additional contributions 
to the distributions beyond the purely classical one discussed here.} in the entire
 kinematic domain of transverse momenta--from the smallest infrared scale given by the 
lattice size  to the largest ultraviolet scale given by the lattice spacing. In physical 
units, these correspond to the inverse nuclear size and large momenta 
($p_T,q_T \gg \qs$). 

One can of course, on dimensional grounds, always express the r.h.s of 
\eq(\ref{eq:C2part2}) as shown. However, $\kappa_2$ in general can depend on 
 the dimensionless ratios $\qs/p_T$, $\qs/q_T$, the relative azimuthal angle
 $\Delta \varphi$ between the two gluons, and the dimensionless combinations  
$\qs \ra$ 
and $m/\qs$, where $m$ is an infrared scale of order $\lqcd$ and $\ra$ 
is the nuclear  radius. We will examine the dependence of $\kappa_2$ on 
these parameters carefully  and discuss what they tell us about the 
structure of correlations. We find that the Glasma flux tube picture
 is valid; however, while the size of transverse correlations is still
 a semi-hard scale, it is not simply $\qs$ and does display some 
sensitivity to infrared physics.  Our results are important for 
quantitative explanations of the ridge as resulting from Glasma
 flux tubes. In the 
semi-quantitative computation of Ref.~\cite{Gavin:2008ev,*Moschelli:2009tg},
 $\kappa_2$ was  implicitly taken to be a free parameter and fit 
to the data--we will compare our  result to this value. We will
 also compare our result to the value obtained by comparing the 
Glasma multiplicity  distribution~\cite{Gelis:2009wh} to PHENIX 
data on multiplicity distributions at RHIC~\cite{Adare:2008ns}.
 
In section \ref{sec:pert}, we will review perturbative 
computations of multi-particle production in the Glasma.  The 
non-perturbative computation 
will be discussed in Section~\ref{sec:nonpert}. We will briefly 
outline the  numerical approach, clearly 
stating the lattice parameters, their relation to physical parameters, 
and the approach to the continuum limit in the transverse and 
longitudinal co-ordinates. We will then systematically study the 
dependence of our  results on the saturation scale, the nuclear 
size and the infrared cutoff $m$. In section \ref{sec:results},
 we will discuss the physical implications of our results and 
further refinements for quantitative comparison with experiment.
We will end with a brief summary of  our key results. Some details 
of the numerical computation are discusssed in an appendix. 

\section{Perturbative results for multi-particle production}
\label{sec:pert}

The perturbative computation of two, three and $n$-gluon correlations 
in the Glasma 
have been discussed elsewhere~\cite{Dumitru:2008wn,Dusling:2009ar,Gelis:2009wh}. 
For completeness, we will briefly review the results here. 
The correlated two particle inclusive distribution can be
expressed as
\begin{equation}
  C(\p,\q) = \frac{1}{4(2\pi)^6}\sum_{a,a'; \lambda,\lambda^\prime} 
    \left< |{\cal M}_{\lambda\lambda^\prime}^{a a'}(\p,\q)|^2\right>
    -
    \left<|{\cal M}_{\lambda}^a(\p)|^2\right>
    \left<|{\cal M}_{\lambda^\prime}^{a'}(\q)|^2\right> \; ,
\label{eq:C2one}
\end{equation}
where the classical contribution to the amplitude for the production
of a pair of gluons with momenta $\p$ and $\q$ is
\begin{eqnarray} 
{\cal M}_{\lambda\lambda^\prime}^{a a'}(\p,\q) &=&
  \epsilon_\mu^\lambda(\p) \,\epsilon_\nu^{\lambda^\prime}(\q) \,p^2q^2\,
  A^{\mu,a}(\p)\,A^{\nu,a'} (\q) \; ,\nonumber\\
  {\cal M}_{\lambda}^{a}(\p)&=& \epsilon_\mu^\lambda(\p)\,p^2\,
A^{\mu,a}(\p)\; .
\label{eq:C2two}
\end{eqnarray}
Here the $\epsilon$'s are the polarization vectors of the gluons and
$a,a^\prime$ are the color indices of the gauge fields.  The average
$\left<\cdots \right>$ in \eq(\ref{eq:C2one}) is an average over the
color configurations of the two nuclei; this average will be discussed
further shortly. 

The gauge fields have a very non-trivial, non--linear  dependence on the sources $\rho_1$ and 
$\rho_2$, which evolves as a function of the proper time $\tau$. As mentioned previously, they can be determined by numerically solving 
Yang-Mills equations for $\tau\geq 0$ with initial conditions at $\tau=0$ given by the gauge fields of each of the nuclei before the 
collision~\cite{Krasnitz:1998ns,Krasnitz:2001qu,Lappi:2003bi}. We will discuss these numerical solutions further in the next section. 
For large transverse momenta $p_T \gg \qs$, however, the equations of motion can be linearized and one can express the 
classical gauge fields produced in
the nuclear collision as~\cite{Kovner:1995ts,Kovner:1995ja,Dumitru:2001ux,Blaizot:2004wu}
\begin{equation}
p^2 A^{\mu,a} (\p) = -i f_{abc}\,g^3\,
\int \frac{\ud^2 \kt}{(2\pi)^2}\, 
L^\mu (\p,\kt)\, 
\frac{\rho_1^b(\kt) \rho_2^c (\pt-\kt)}{\kt^2 (\pt-\kt)^2} \; .
\label{eq:C2three}
\end{equation}
Here $f_{abc}$ are the SU(3) structure constants, $L^\mu$ is the
well known\footnote{The components of this four vector are given
  explicitly by $L^+(\p,\kt) = -\frac{\kt^2}{p^-}$, $L^-(\p,\kt) =
  \frac{(\pt-\kt)^2-\pt^2}{p^+}$,
  $L^i(\p,\kt) = -2\, \kt^i$.} Lipatov
vertex and $\rho_1,\rho_2$ are
respectively the Fourier transforms of the color charge densities in
the two nuclei~\cite{Iancu:2003xm}. 

From \eq(\ref{eq:ngluon-LLog}), the average in \eq(\ref{eq:C2one}) corresponds to
\begin{equation}
\left< {\cal O} \right> \equiv \int [D\rho_1\,D\rho_2] 
Z[\rho_1] Z[\rho_2] \,{\cal O}[\rho_1,\rho_2] \; . 
\label{eq:avg}
\end{equation}
In the MV model~\cite{McLerran:1994ni,*McLerran:1994ka,*McLerran:1994vd,Kovchegov:1996ty}
\begin{equation}
  Z[\rho] \equiv \exp\left(-\int \ud^2
  \xt \frac{\rho^a(\xt)
    \rho^a(\xt)}{2\,\mu_{_A}^2}\right)\; ,
\label{eq:avg1}
\end{equation}
where $\rho$ can be either $\rho_1$ or $\rho_2$. As we will discuss further in the 
next section, the color charge squared per unit area
$\mu_{_A}^2$ can be expressed simply in terms of $\qs$. We will consider
this Gaussian model in the rest of this paper\footnote{In the simplest
  treatment of small $x$ evolution, based on the Balitsky-Kovchegov
  equation~\cite{Balitsky:1995ub,*Balitsky:1998kc,*Balitsky:1998ya,*Kovchegov:1999yj,*Kovchegov:1999ua,*Balitsky:2001re}, $Z[\rho]$ can also be modelled by a
  Gaussian, albeit one with a non-local variance~\cite{Blaizot:2004wu,Fujii:2006ab}.}. For these Gaussian correlations,
in momentum space,
\begin{equation}
\left< \rho^a(\kt) \rho^b (\kt^\prime)\right> =
(2\pi)^2 \mu_{_A}^2 \,\delta^{ab} \delta(\kt-\kt^\prime)\; .
\end{equation}

Using \eq\nr{eq:C2three}, \nr{eq:avg} and \nr{eq:avg1} in \eq\nr{eq:C2one}, 
one obtains~\cite{Dumitru:2008wn},
\begin{equation}
C(\p,\q) 
= 
\frac{S_\perp}{16\,\pi^7}\, 
\frac{(g^2\mu_{_A})^8}{g^4\, \qs^2}\, 
\frac{ N_c^2 (N_c^2-1)}{p_T^4 \,q_T^4} \; .
\label{eq:C2-final}
\end{equation}
It is instructive to express the
result in \eq\nr{eq:C2-final} in terms of the inclusive single
gluon spectrum. This result, due originally to Gunion and Bertsch~\cite{Gunion:1981qs}, 
has been been recovered previously in the CGC
framework~\cite{Kovner:1995ts,Kovner:1995ja,Kovchegov:1997ke,Gyulassy:1997vt} 
and is known to have the form
\begin{equation}
\left< \frac{\ud N}{\ud y_p \ud^2 \pt}\right> 
= 
\frac{S_\perp}{4 \pi^4}\, 
\frac{(g^2\mu_{_A})^4}{g^2}\, 
\frac{N_c (N_c^2-1)}{p_T^4}\, \ln\left(\frac{p_T}{\qs}\right) \; .
\label{eq:single-gluon}
\end{equation}

Substituting \eq\nr{eq:single-gluon} on the r.h.s of \eq\nr{eq:C2-final}, one obtains 
\begin{equation}
 C(\p,\q) 
=  
\frac{\kappa_2}{S_\perp \qs^2} 
\left< \frac{\ud N}{\ud y_p \ud^2\pt}\right>  
\left< \frac{\ud N}{\ud y_q \ud^2\qt}\right> \; ,
\label{eq:k2-pert}
\end{equation} 
which is the result we noted in \eq(\ref{eq:C2part2}), with\footnote{In the computation of Ref.~\cite{Dumitru:2008wn}, there were numerical errors 
which gave a significantly larger perturbative estimate for $\kappa_2$.} 
\begin{equation}
\kappa_2 \approx  \frac{\pi}{(N_c^2-1)} = 0.4
\label{eq:k2-pert-num}
\end{equation}
Identifying the theoretical error on $\kappa_2$ is difficult at this 
stage from the perturbative computation. 

Computing $\kappa_2$ non--perturbatively by numerically solving the Yang--Mills equations 
is the primary objective of this paper. This quantity is not a pure number but can 
contain interesting structure. 
In the perturbative calculation, $\kappa_2$ is independent of the relative angle 
$\Delta \varphi$ at very large 
transverse momenta $p_T$, $q_T \gg \qs$. However, even in the perturbative 
calculations, one notices that there are finite azimuthal correlations between gluons 
as one goes away from the limit of asymptotically large transverse momenta. It is 
important to understand these correlations at momenta of interest to experiment to 
ascertain whether they have any phenomenological significance. Further, $\kappa_2$ 
can depend non-trivially on the transverse 
momenta of the pairs and on the energy and centrality of the nuclear collision. 
Finally, we want to determine whether $\kappa_2$ is infrared finite. 
This was the case for the analogous factor for the non-perturbative single gluon 
distribution at any finite time~\cite{Krasnitz:1998ns}; there is no guarantee that
 this should be the case for multi-gluon distributions. 

Before we end this section, we should mention that perturbative
 computations were 
also performed for 3-gluon~\cite{Dusling:2009ar} and $n$-gluon 
correlations~\cite{Gelis:2009wh}. The result can be nicely 
summarized in terms of the cumulants of the multiplicity distribution
 as~\cite{Gelis:2009wh}
\begin{equation}
\left< \frac{\ud^n N}{\ud y_1 \ud^2 \pt_1 \cdots \ud y_n \ud^2 \pt_n}\right> = 
\frac{(n-1)!}{k^{n-1}}  \left< \frac{\ud N}{\ud y_1 \ud^2 \pt_1}\right> 
\times
\cdots 
\times
\left< \frac{\ud N}{\ud y_n \ud^2 \pt_n }\right> \, ,
\label{eq:n-cumulant}
\end{equation}
where 
\begin{equation}
k = \zeta \, {(N_c^2 -1) \qs^2 S_\perp \over 2\pi} \, .
\label{eq:n-cum-k}
\end{equation}
Here $\zeta$, is a non--perturbative coefficient\footnote{In the perturbative 
computation for the two and three particle  correlations, this number 
was taken to be $\zeta =2$.} to be determined by numerical solutions of 
Yang-Mills equations. The expression in \eq(\ref{eq:n-cumulant}) corresponds 
to a negative binomial distribution. One can extract $\zeta$ 
from fits to the multiplicity distribution data~\cite{Adare:2008ns} and compare our 
non-perturbative result for $\kappa_2$ to this value thereby testing the 
validity of \eq(\ref{eq:n-cumulant}) for the 2nd cumulant. Our approach 
can be straightforwardly be extended to higher cumulants but the computations are numerically intensive and will not be considered further here.

\section{Non-perturbative computation}
\label{sec:nonpert}

The two gluon cumulant, on dimensional grounds, can always be expressed in the 
form given in \eq(\ref{eq:k2-pert}). However, as we have noted 
previously, the dimensionless coefficient $\kappa_2$ can in general be a function of  $\qs/p_T, \qs/q_T, \Delta \varphi, \qs \ra$ and $m/\qs$, where 
$m$ is an infrared regulator scale that can be added to the MV model. 
If we assume that $m\sim \lqcd$, the high energy asymptotics of our formalism
is $m/\qs \ll 1$ and $\qs \ra \rightarrow \infty$. 
For A+A collisions at realistic 
RHIC and LHC energies, one expects $m/\qs \sim 0.2-0.1$ and 
$\qs \ra = 35-70$ respectively. We will return to the discussion of 
the infrared scale $m$ later at the end of \se\ref{sec:results}.

For asymptotically large $p_T$, $q_T \gg \qs$, we anticipate, on the 
basis of our perturbative results, that $\kappa_2\rightarrow$ constant. For 
$p_T$, $q_T \leq \qs$, we will determine the dependence of $\kappa_2$ 
on these ratios. In particular, we would like to determine whether the double 
inclusive gluon distribution is rendered infrared safe by the non-linearities 
of the Yang-Mills fields that may generate a mass scale (analogous to a plasmon mass)
 for finite times $\tau \gtrsim 1/\qs$. 

\subsection{Numerical approach} 
\label{sec:numerics}

The numerical solutions of the classical Yang--Mills equations have been discussed 
at length elsewhere~\cite{Krasnitz:1998ns,Krasnitz:2001qu,Lappi:2003bi} and we 
will only outline the approach followed here for completeness. In the 
MV model~\cite{McLerran:1994ni}, the Yang-Mills equations
\begin{equation}\label{eq:ym}
[D_{\mu},F^{\mu \nu}] = J^{\nu}.
\end{equation}
have the source currents 
\begin{equation}\label{eq:twonucl}
 J^{\mu} =  \delta^{\mu +}\rho_{1}(\xt,x^-) 
+ \delta^{\mu -}\rho_{2}(\xt,x^+),
\end{equation}
where the color charge densities $\rho_{1,2}$ of the two nuclei
are independent static color sources localized on the light
cone $\rho_{1,2}(\xt,x^\mp) \sim \rho_{1,2}(\xt)\delta(x^\mp)$. They 
are distributed with the Gaussian 
probability distribution
\begin{equation}\label{eq:korr}
\langle \rho^a(\xt) \rho^b(\yt) \rangle = g^2 \mu_A^2 \delta^{ab}\delta^2(\xt-\yt),
\end{equation}
where $\xt$ and $\yt$ are vectors in the transverse plane. 
The initial conditions for the solutions of these equations 
in the forward light cone in the Fock--Schwinger gauge ($A^\tau =0$) are given by
\begin{eqnarray}\label{eq:initcond}
A^i|_{\tau=0} &=& A^i_{1} + A^i_{2}, \\ \nonumber
A^\eta|_{\tau=0} &=& \frac{ig}{2}[A^i_{1},A^i_{2}]\,,
\end{eqnarray}
with 
\begin{equation} \label{eq:ainit}
A_{1,2}^i = - \frac{i}{g} V_{1,2} \partial^i V_{1,2}^\dagger \,\textrm{ and}\,\,
\nabt^2 \Lambda_{1,2} = -g \rho_{1,2}\,.
\end{equation}
where 
\begin{equation}
V_{1,2} = {\cal P}_\mp \, e^{ i\Lambda_{1,2}}.
\label{eq:Wilson-line}
\end{equation}
It is convenient in the forward light cone to use the co-ordinate system $(\tau,\eta,\xt)$,
 where $\tau^2 = t^2 - z^2$ and 
$\eta = 0.5 \ln((t+z)/(t-z))$. The Yang-Mills equations in this co-ordinate system are manifestly 
boost invariant with the gauge fields 
independent of $\eta$, namely, $A^\mu\equiv A^\mu(\tau,\xt)$. The symbol ${\cal P}_\mp$ 
denotes path ordering in the  $\mp$ directions 
of the Wilson line $V_{1,2}$ respectively, which for nucleus 1 (nucleus 2) have an implicit
 dependence on $x^-$ ($x^+$) in the 
solution. We will return to this point at the end of the subsection.

A subtle point we would like to emphasize is that the Wilson lines $V$, in the gauge 
field solution given by \eq(\ref{eq:ainit}) for nucleus 1 (nucleus 2) before the 
collision, are path ordered in $x^-$ ($x^+$). This feature of the continuum solution 
is implemented by constructing the Wilson lines in consecutive steps representing
a discretization of the longitudinal coordinate. 
On each lattice site $\xt$ one constructs random color charges  with a 
local Gaussian distribution
\begin{equation}\label{eq:discrsrc}
\left\langle \rho^a_k(\xt) \rho^b_l(\yt) \right\rangle =
 \delta^{ab}  \delta^{kl}  \delta^2(\xt-\yt)
\frac{g^2 \mu_A^2}{N_y}\,,
\end{equation}
with the indices $k,l=1,\dots,N_y$ representing a discretized longitudinal coordinate.
Numerical calculations of the single inclusive gluon distributions previously used $N_y=1$,
 whereas the derivation
of the analytical expression in \eq(\ref{eq:ainit}) required an extended source corresponding 
to the limit $N_y \to \infty$.

Our normalization is chosen to give 
\begin{equation}
\sum_{k,l} \left\langle \rho^a_k(\xt) \rho^b_l(\yt) \right\rangle =
 \delta^{ab}    \delta^2(\xt-\yt)
g^2 \mu_A^2.
\end{equation}
The Wilson lines for the initial condition are then constructed from the sources 
\eq(\ref{eq:discrsrc}) by solving a 
Poisson equation and exponentiating it to give 
\begin{equation}\label{eq:uprod}
V_{1,2}(\xt)  = \prod_{k=1}^{N_y} \exp\left\{ -i g \frac{\rho_k^{1,2}(\xt)}{\nabt^2 + m^2}\right\}.
\end{equation}
Here we have introduced an additional infrared regulator $m$ into
 the MV model; some kind of 
infrared cutoff is required for inverting the Laplace operator. 
The single inclusive gluon distribution, integrated over transverse momenta, 
turns out to be weakly dependent on the infrared cutoff. One of the central 
aims of this paper is to study whether this is also the case
for correlations.
The parameter $m$ plays an important role in the interpretation of our results and we will return to it in our discussion of the results of our computations.
We will return to the physical interpretation of the role of the parameter $m$ later in our 
discussion. For large $N_y$ the charge densities $\rho_k$ in \eq(\ref{eq:discrsrc}) become
small, and the individual elements in the product \nr{eq:uprod} approach identity.
This is precisely the procedure that leads  in the $N_y \to  \infty$ limit to
the continuum  path ordered exponential in \eq(\ref{eq:Wilson-line}). 

The previous numerical computations of the single inclusive spectrum defined the gluon multiplicity in a manner that slightly deviates
from the definition based on the LSZ formalism in  \eq\nr{eq:AA}.
In \cite{Lappi:2003bi}, taking advantage of the equipartition of energy in the classical 
theory, only the electric field 
parts of the numerical solution were used. Assuming a 
free dispersion relation the gluon multiplicity was taken to be
\begin{equation}\label{eq:defmultiee}
\frac{\ud N}{\ud^2 \kt \ud y} = \frac{1}{(2 \pi)^2}\frac{2}{|\kt|} 
\Tr \bigg[
\frac{1}{\tau} E^i(\kt)E^i(-\kt) 
+\tau E^{\eta}(\kt) E^{\eta}(-\kt)
\bigg],
\end{equation}
where the electric fields
$E^i = \tau \dot{A}_i$ and $E^{\eta} = \dot{A}_\eta/\tau$ are time derivatives of the gauge 
potentials, defined here with the explicit factors of $\tau$ coming from the formulation 
of the theory in $\tau,\eta$-coordinates.
In Refs.~\cite{Krasnitz:1998ns,Krasnitz:2001qu}
 the numerically obtained $E$- and $A$-fields were used to determine the dispersion relation of the interacting theory. This was then in turn used 
to determine the multiplicity, unlike the free dispersion relation assumed in 
\eq\nr{eq:AA}. In this work
we use a definition that corresponds to 
\eq\nr{eq:AA} and take the gluon spectrum as
\begin{align}\label{eq:defmulti}
\frac{\ud N}{\ud^2 \kt \ud y} = \frac{1}{(2 \pi)^2} \Tr \bigg\{ &
\frac{1}{\tau |\kt|} E^i(\kt)E^i(-\kt) 
+ \tau |\kt|  A_i(\kt)A_i(-\kt) 
\\ &
+ \frac{1}{|\kt|} \tau E^\eta(\kt) E^\eta(-\kt)
+ \frac{|\kt|}{\tau} A_\eta(\kt) A_\eta(-\kt)
\\ & \label{eq:multitrint}
+ i \Big[ E^i(\kt) A_i(-\kt) - A_i(\kt)E^i(-\kt) \Big]
\\ & \label{eq:multietaint}
 +i \Big[ E^\eta(\kt) A_\eta(-\kt) -  A_\eta(\kt) E^\eta(-\kt) \Big]
\bigg\},
\end{align}
where the fields are evaluated in the 2-dimensional Coulomb gauge.
The interference terms \nr{eq:multitrint} and \nr{eq:multietaint} are odd under the 
transformation $\kt \to -\kt$. They do not contribute when the gluon spectrum is
averaged over the angle of $\kt$, integrated over $\kt$ or averaged over configurations
of the sources. Thus neglecting them was justified when one was interested in the single
inclusive gluon spectrum. In the case of two-gluon correlations they cannot, however, be
 neglected\footnote{We thank F. Gelis for pointing this out to us.}.  Due to these interference terms, the symmetry $n(\kt) = n(-\kt)$ does not 
 hold configuration by configuration, but only on the average. Thus the correlation function
$C(\pt,\qt) \neq C(\pt,-\qt)$. In particular, there is a peak in the correlation at
$\pt=\qt$, which without the interference terms \nr{eq:multitrint} and \nr{eq:multietaint}
would, by symmetry, imply a similar peak at $\pt=-\qt$. The main numerical 
effect of including these terms is that the backward peak at $\pt=-\qt$ is significantly 
depleted.

\subsection{Parameters in the computation}
\label{sec:param}

From the discussion in the previous subsection, the parameters in the numerical 
lattice computation are
\begin{itemize}
\item $g^2 \mu_A$, the root mean square color charge density. 
\item $N_y$, the number of points in the discretization of the longitudinal ($x^-$ 
or rapidity) direction. 
\item The infrared regulator $m$.
When $m=0$, the Poisson equation is solved by leaving out the zero transverse
 momentum mode. This procedure 
corresponds to an infrared cutoff given by the size of the system.
\item The lattice spacing $a$.
\item The number of transverse lattice sites $N_T$, giving the transverse 
size of the lattice $L = N_T a$.
\end{itemize}
Of the dimensionful parameters $a$, $g^2\mu_A$ and $m$, only the dimensionless 
combinations $g^2\mu_A a$ and $m a$ appear in the numerical calculation. The continuum 
limit $a\to 0$ is taken by letting $N_T \to \infty$ such that
$g^2 \mu_A a \to 0$ and $g^2 \mu_A L = g^2 \mu_A a N_T$ is a constant. 
This constant is determined by the physics input of the calculation. 
We have $\pi \ra^2 = L^2$ and, as we shall discuss in the next subsection, $g^2\mu_A$ 
is simply related to the saturation scale $\qs$. The physical combination $\qs \ra$ 
relevant at RHIC and LHC energies will translate into corresponding values of 
$g^2\mu L$ in our computation. In \se\ref{sec:results}, 
we will explicitly translate our numerical results into physically relevant numbers. 
We note that there is another dimensionless combination $m/g^2\mu_A$-we will study the
sensitivity of our results to this ratio.

\subsection{Relating $g^2\mu_A$  and $\qs$ on the lattice}

The root mean squared color charge density $g^2 \mu_A$ that appears in our Glasma computations can be simply related to the 
saturation scale $\qs$ extracted from deeply inelastic scattering experiments. In these experiments, the cross section for a virtual photon 
scattering off a high energy hadron or nucleus is simply related to the dipole cross section of a quark-antiquark pair scattering off the hadron. This 
``dipole'' cross-section is determined, in the CGC formalism, by the correlator of two Wilson lines in the fundamental 
representation of each of the nuclei before the collision as
\begin{equation} \label{eq:wlinecorr}
\widetilde{C}^F(\xt-\yt) = \langle  \trace V^\dag(\xt) V(\yt)\rangle\,,
\end{equation}
with the expectation value $\langle \rangle$ evaluated with the distribution 
of the sources. For Gaussian correlators in the MV model of sources extended 
in the longitudinal direction, 
\begin{equation}
\langle \rho^a(\xt,x^-)\rho^b(\yt,y^-)\rangle = g^2 \delta^{ab} \delta^2(\xt -\yt)\delta(x^- - y^-)\nonumber \mu_A^2(x^-) ,
\label{eq:ext-sources}
\end{equation}
the Wilson line correlators can be computed
analytically up to a logarithmic
infrared cutoff that must be introduced in solving the Poisson equation in \eq(\ref{eq:ainit}).
The result is 
\begin{equation} \label{eq:extcorrsF}
\widetilde{C}^F(\xt) \approx N_c \,\exp\left({\frac{C_F}{8 \pi} \chi \xt^2 \ln(m |\xt |)}\right) \,, 
\end{equation}
with
\begin{equation} \label{eq:extmv}
\chi = g^4 \int dx^- \mu_A^2(x^-).
\end{equation}
Alternately, the saturation scale $\tilde{\qs}$, in the fundamental representation,
is defined as the momentum corresponding to the maximum of $\kt^2 {\tilde C}^F (\kt)$,
 where  ${\tilde C}^F(\kt)$ is the Fourier transform of \eq(\ref{eq:wlinecorr}). In 
this manner, one can relate the saturation scale to the root mean square color 
charge density.

\FIGURE{
\includegraphics[width=0.45\textwidth]{adjuspectscqs.eps}
\caption{
Adjoint representation Wilson line correlator for different 
$N_y$ as a function of $\kt/\qs$.  From Ref.~\cite{Lappi:2007ku}. 
}\label{fig:wlinecorr}
}
The saturation scale defined previously is an inverse correlation length associated with the correlator of two Wilson lines in the fundamental representation. 
In a nuclear collision, it is the saturation scale in the adjoint representation that is relevant. It is defined, equivalently as the momentum corresponding 
to the maximum of $\kt^2 {\tilde C}^A(\kt)$, where ${\tilde C}^A(\kt)$ 
is the Fourier transform of the correlator of two adjoint 
Wilson lines
\begin{equation} \label{eq:adjwlinecorr}
C^A(\xt) = \langle V_{ab}(\xt) V_{ab}(0) \rangle\,,
\end{equation}
with 
\begin{equation} \label{eq:adjU}
V_{ab}(\xt) = 2\trace \left[t^a V^\dag(\xt) t^b V(\xt) \right]\,.
\end{equation}
With some algebra, in the large $N_c$ limit, the adjoint representation correlator can be related
to a higher correlator of fundamental representation Wilson lines
\begin{equation} \label{eq:adjwlinecorr2}
C^A(\xt) =  \left\langle  \left| \trace \left[V^\dag(\xt)V(0) \right] 
 \right|^2 
- 1 \right\rangle \, .
\end{equation}
In the Gaussian MV model, one obtains
\begin{equation}
C^A(\xt) \approx (N_c^2 -1)\,\exp\left({\frac{N_c}{8 \pi} \chi \xt^2 
\ln(m |\xt|)}\right)\,.
\end{equation}
The saturation scale $\qs$ in the adjoint representation, in identical fashion 
to the fundamental case, is defined as the momentum corresponding to the maximum of $\kt^2 {\tilde C}^A (\kt)$, where ${\tilde C}^A(\kt)$ is the Fourier transform of
 \eq\nr{eq:adjwlinecorr2}. The adjoint Wilson line correlator is plotted in 
\fig\ref{fig:wlinecorr}
The two saturation scales are approximately related by the ratios of the  Casimirs of the representations, namely, $\qs^2 \approx \frac{C_A}{C_F}\tilde{\qs}^2$.

\FIGURE{
\includegraphics[width=0.45\textwidth,clip=true]{ny.eps}
\includegraphics[width=0.45\textwidth,clip=true]{g2muvsm.eps}
\caption{
Left: $\qs/g^2\mu_A$ versus $N_y$. From Ref.~\cite{Lappi:2007ku}. 
Right: Dependence of the ratio $\qs/g^2\mu$ on $m/\qs$ for a fixed 
$\qs= 1\gev$ and
$N_y = 50$. These are the conversion factors used to obtain the
$m$-dependence of our results for fixed $\qs$ in 
\fig\protect\ref{fig:m_fixedqs}.
}\label{fig:qsg2muvsny}\label{fig:qsg2muvsm}
}
Our definition of the saturation scale is not sensitive to the exact shape of the correlator for very large
or small transverse momenta, and for a Gaussian correlator it reproduces
the GBW saturation scale as $1/R^2_0 = \tilde{\qs}^2$.
For small momenta, the fundamental and adjoint correlators look like Gaussians, which is the form 
used in the ``GBW'' fit of DIS data in 
Refs.~\cite{Golec-Biernat:1998js,*Golec-Biernat:1999qd,*Stasto:2000er}.
For large momenta there is a power law tail $1/\kt^4$ that differs from the
original GBW fits, but resembles more closely the form required to match smoothly
to DGLAP evolution for large $Q^2$ \cite{Bartels:2002cj}.

In Ref.~\cite{Lappi:2007ku}, the relation between $\qs$ and $g^2\mu_A$ was studied 
extensively employing numerical lattice techniques.  Discretizing the longitudinal 
distribution of 
sources as described in \eq(\ref{eq:discrsrc}) and constructing the Wilson line as
 in \eq(\ref{eq:uprod}), it was shown that $\qs \sim 0.57 g^2 \mu_A$ for 
$N_y=1$. This value was used in the numerical Glasma simulations 
of the single inclusive gluon spectrum.  
As we shall see in the following the gluon spectrum, when expressed in terms of the
scaling variable $p_T/\qs$, is to a very good approximation independent of $N_y$. This 
is however not the case for the double inclusive spectrum, which will have a stronger
dependence on $N_y$. 

Due to the $m$-dependence of the single inclusive spectrum, the
ratio $\kappa_2$ defined by \eq\nr{eq:C2part2} will depend on $m$. To study this
 $m$-dependence for larger values of $N_y$ (closer to the continuum limit in the
longitudinal direction), we will need to invert the computation of $\qs$ as a function 
of $g^2\mu$ to obtain, as a function of $m/\qs$, the required values of $g^2\mu$ 
corresponding to a fixed value of $\qs$. The result of this exercise for 
$\qs=1\gev$ and $3\gev$ and using the numerical method employed in ~\cite{Lappi:2007ku} 
is shown in \fig\ref{fig:qsg2muvsm}.

\section{Results}
\label{sec:results} 
We begin the discussion of our numerical results with the single particle
 spectrum because we are interested in the ratio of the double inclusive spectrum 
to the single particle spectrum squared. Some of the results for the single particle spectrum are 
new and have not been published elsewhere. We will then proceed to discuss the 
double inclusive spectrum and state our results for $\kappa_2$ defined in 
\eq\nr{eq:C2part2}. 
\subsection{Single inclusive spectrum}
The single inclusive multiplicity in A+A collisions was computed extensively previously by solving Yang-Mills equations numerically. 
We note that the single inclusive multiplicity computed gives 
excellent fits to the RHIC multiplicity data~\cite{Krasnitz:2003jw,Lappi:2007ku} with values of $\qs$ for gold nuclei that agree to $\sim 10$\% with those 
extracted from extrapolations of HERA data to RHIC~\cite{Kowalski:2007rw}. In this sub-section, we will address some details of the computation 
that have not been presented previously and are relevant for the computation of $\kappa_2$. 

In Ref.~\cite{Lappi:2007ku}, the dependence of the single inclusive gluon 
distribution as a function of $N_y$ was not computed.
We have done it here and the result is shown in \fig\ref{fig:single-inc-Nydep}.
 It shows that the single inclusive spectrum, scaled in 
units of $(p_T/\qs)^2$, shows virtually no $N_y$ dependence in particular for 
the moderate $p_T$ region that dominates the integrated multiplicity.
We may conclude  that it has a weak dependence on $N_y$ (for fixed $m/\qs$)
 addressing a concern  raised in Ref.~\cite{Fukushima:2007ki}. 
The numerical single particle spectrum in the UV region 
is sensitive to the lattice spacing $a$, which, for a fixed value of 
$\qs \ra$, translates into a dependence on the size of the lattice $N_T$.
As shown in \fig\ref{fig:single-inc-Ndep}, the spectrum approaches the 
continuum ultraviolet behavior of $\qs^4/p_T^4$ with increasing 
lattice sizes from $N_T = 128^2$ to $N_T=512^2$ lattices. 

\FIGURE{
\includegraphics[angle=-90,width=0.45\textwidth]{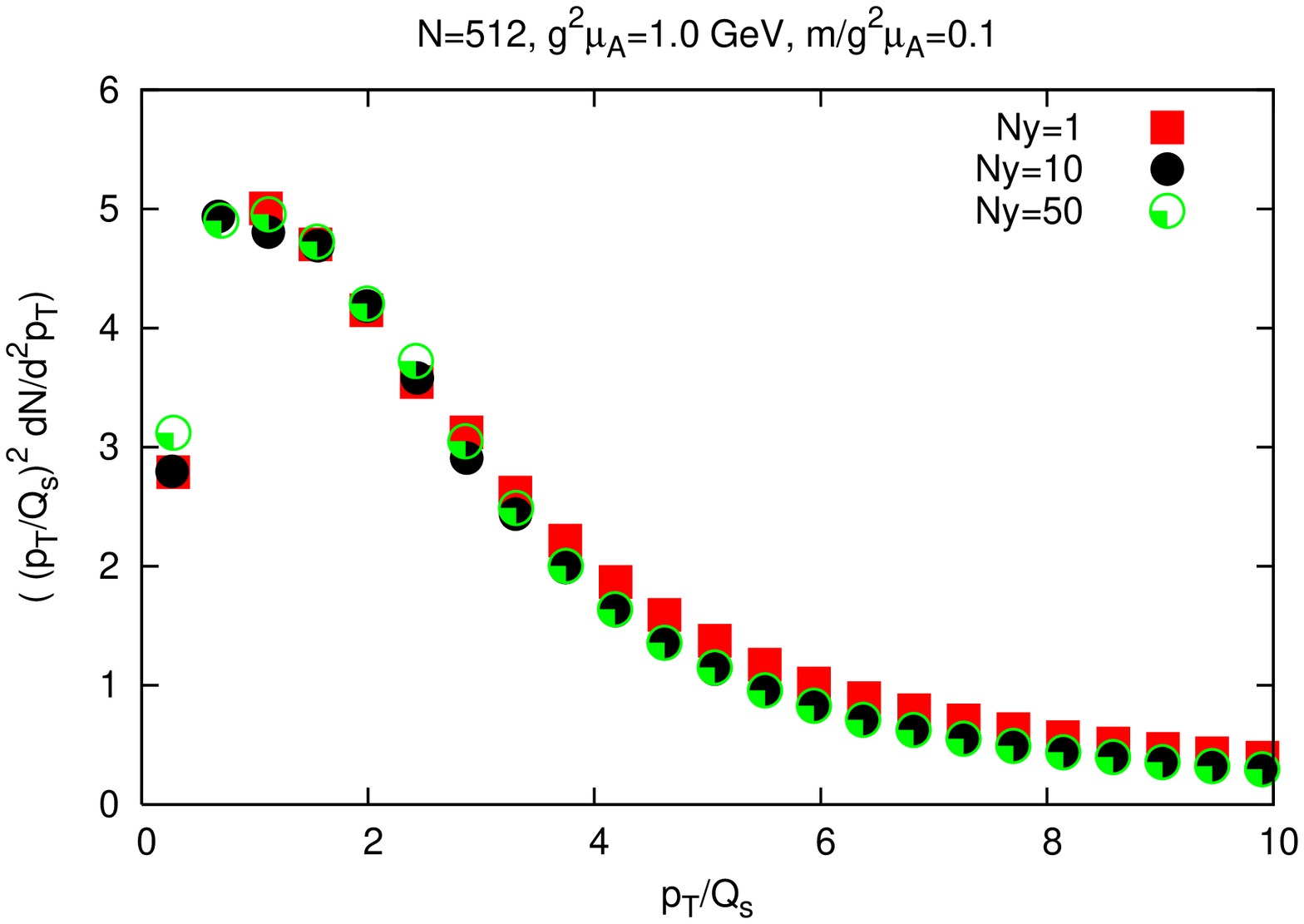}
\includegraphics[angle=-90,width=0.45\textwidth]{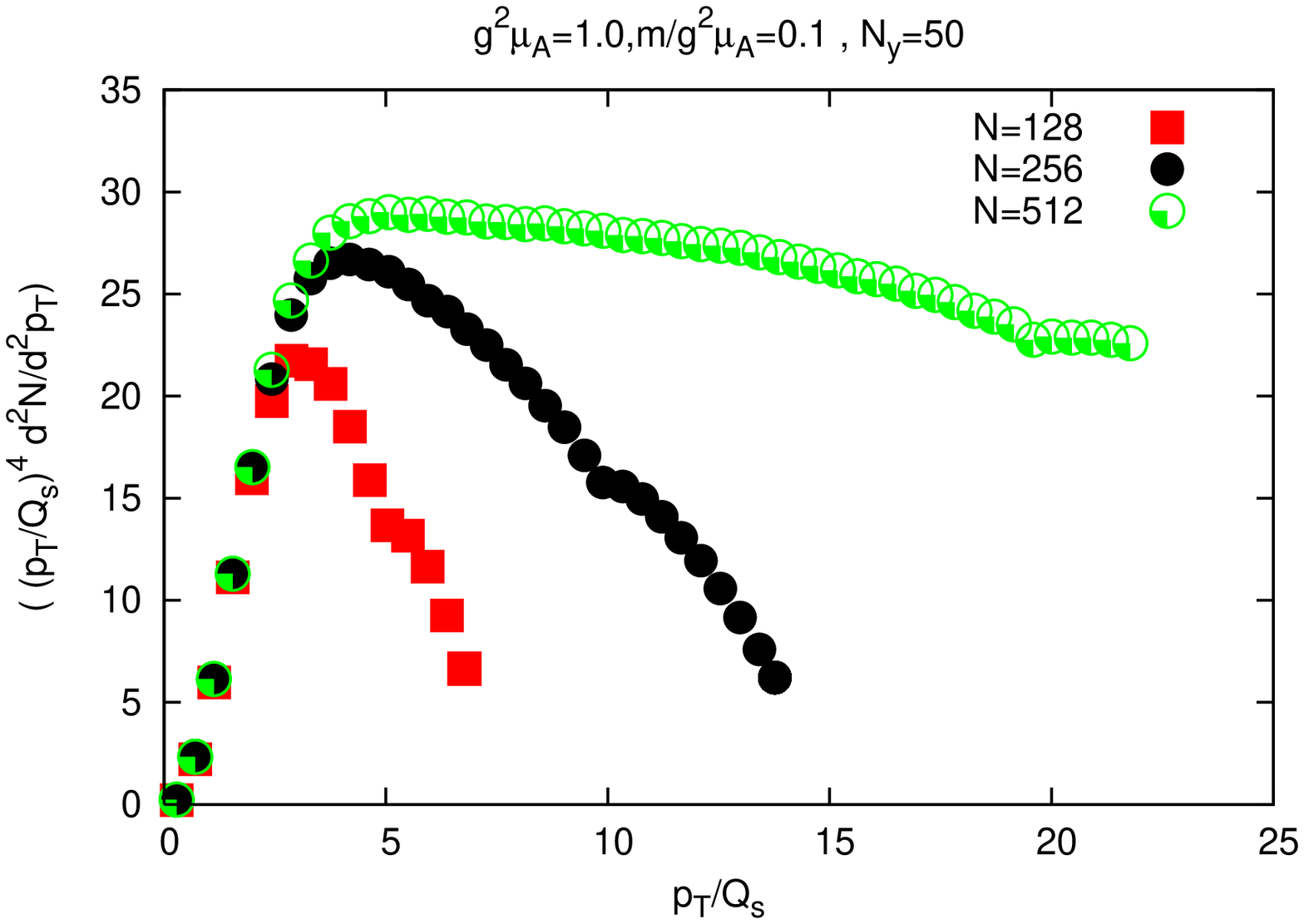}
\caption{
Left: The single particle spectrum (scaled by $(p_T/\qs)^2$) for 
different $N_y$. 
Parameters are $g^2{\mu}=1\gev$, $N_T=512$, IR cutoff fixed at $0.1g^2{\mu}$. 
Right: The single particle spectrum (scaled by $(p_T/\qs)^4$) for different 
$N_T$. 
The continuum spectrum has a $\qs^4/p_T^4$ behavior at $p_T\gg \qs$; 
the convergence to this continuum behavior is clearly seen with increasing lattice size.
}
\label{fig:single-inc-Nydep}\label{fig:single-inc-Ndep}
}

We studied the $m/\qs$ dependence of the single inclusive spectrum for a fixed 
large $N_y=50$ and a $512\times 512$ transverse lattice. The result is shown in 
\fig\ref{fig:single-inc-mdep} (left). For small $m/\qs$, the dependence of the spectrum on this quantity is quite weak-changing $m/\qs$ by a factor 
of 2 shows virtually no change in the spectrum. However, some dependence is seen when $m/\qs$ is increased, albeit the dependence is 
relatively weak in the $p_T\sim \qs$ region which gives the dominant contribution to the integrated multiplicity. It should be kept in mind that 
the dependence of the spectrum on $m/\qs$ as shown is weaker than a logarithmic dependence even in the region where the dependence is the largest. 
It might seem counterintuitive that the value of an infrared scale $m$ affects the gluon
spectrum at such high momenta. One way to understand this is to keep in mind that 
the unitarity of the Wilson lines imposes a sum rule on their correlators, 
\eqs\nr{eq:wlinecorr} and~\nr{eq:adjwlinecorr2}. Namely, because
$\widetilde{C}^F(\xt-\yt=0)=\nc$, the integral of the momentum space correlator
$\int\ud^2 \kt \widetilde{C}^F(\kt) = (2\pi)^2\nc$ is a constant. 
Thus a modification of the distribution in
the infrared will also affect the UV behavior, which is also reflected in the
gluon spectrum.

Finally, we plot in \fig\ref{fig:single-inc-g2mudep} (right) the dependence of the single particle spectrum on 
the other dimensionless combination of scales $g^2\mu L$
for fixed $m/g^2\mu$. (This corresponds to a very good approximation to fixed 
$m/\qs$.). Virtually no dependence is seen on this quantity, confirming the 
expectation that the single inclusive spectrum is completely insensitive to physics on scales corresponding to the size of the system. In summary therefore, 
the single particle spectrum is most sensitive to physics at the scale $\qs$, weakly sensitive to physics on the scale $m$ and completely insensitive to 
physics on the scale $1/R$. 
\FIGURE{
\includegraphics[angle=-90,width=0.45\textwidth]{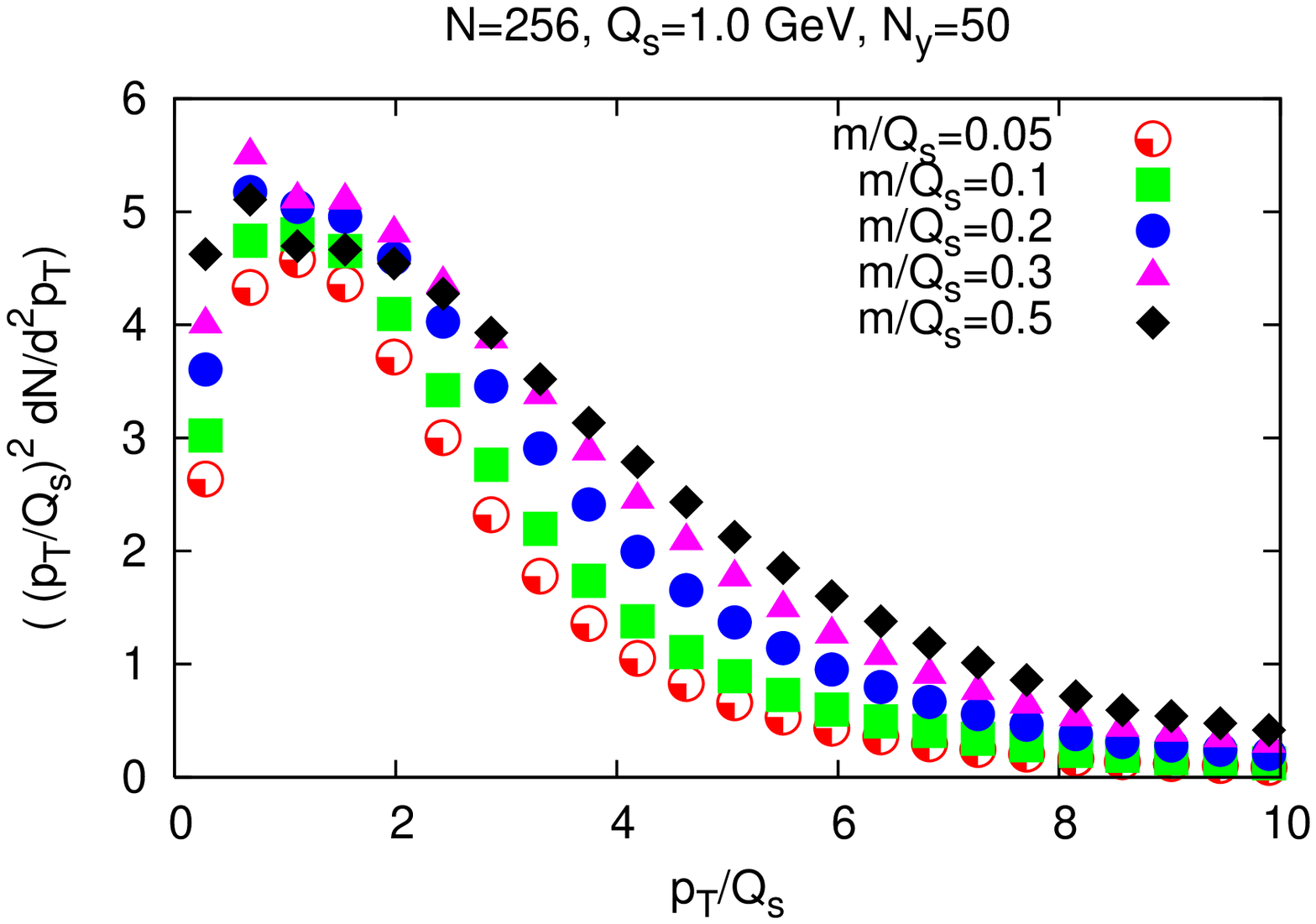}
\includegraphics[angle=-90,width=0.45\textwidth]{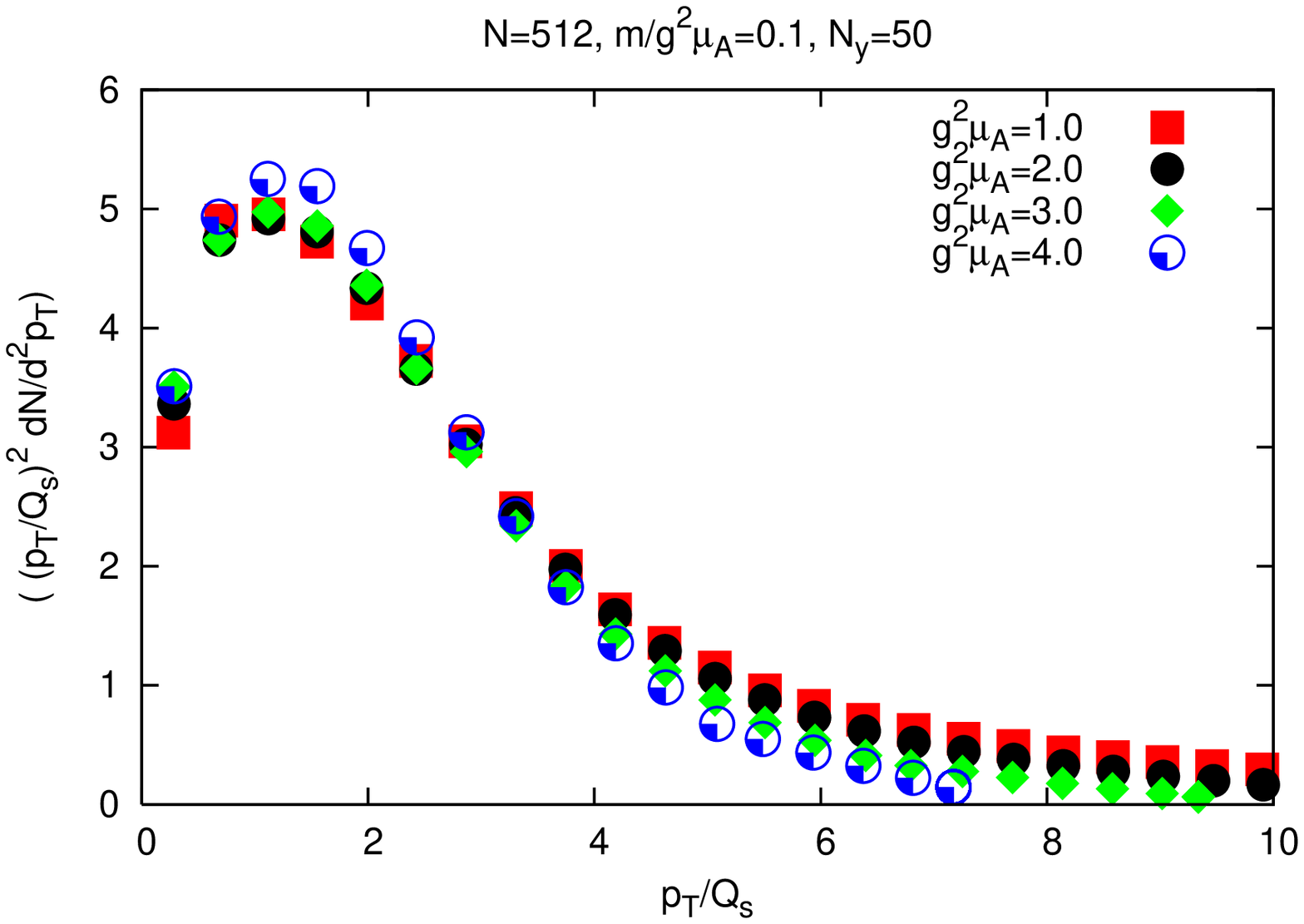}
\caption{
Left: The single particle spectrum for different $m$ for $N_y=50$, $N_T=512$, 
and $\qs=1\gev$.
 The spectrum is again scaled by $(p_T/\qs)^2$.
 Right: The dependence of the single particle spectrum on $g^2{\mu}$ for a 
fixed $m/g^2\mu$. 
}\label{fig:single-inc-g2mudep}
\label{fig:single-inc-mdep}
}

\subsection{Double inclusive spectrum}

\FIGURE{
\includegraphics[angle=-90,width=0.45\textwidth]{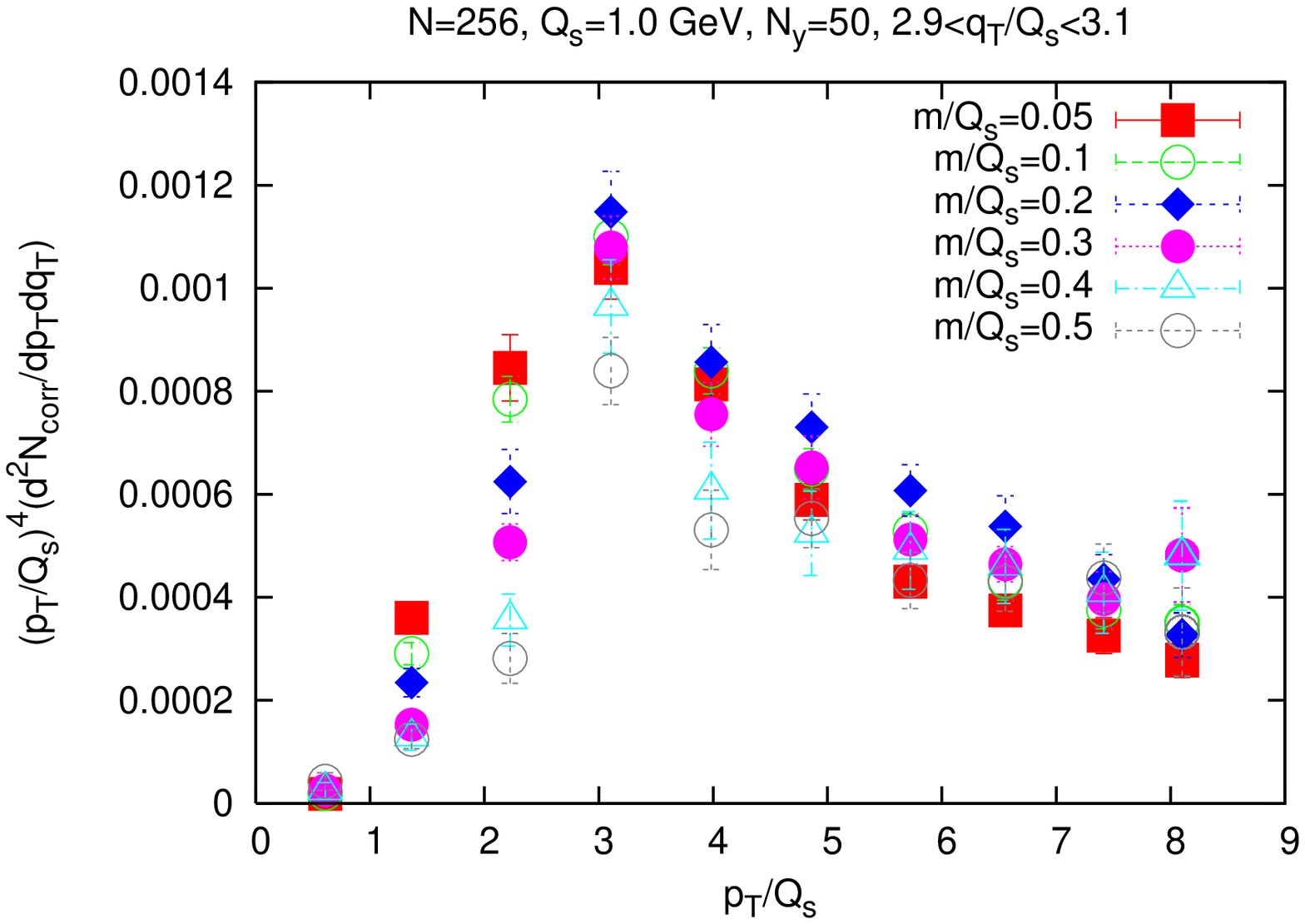}
\includegraphics[angle=-90,width=0.45\textwidth]{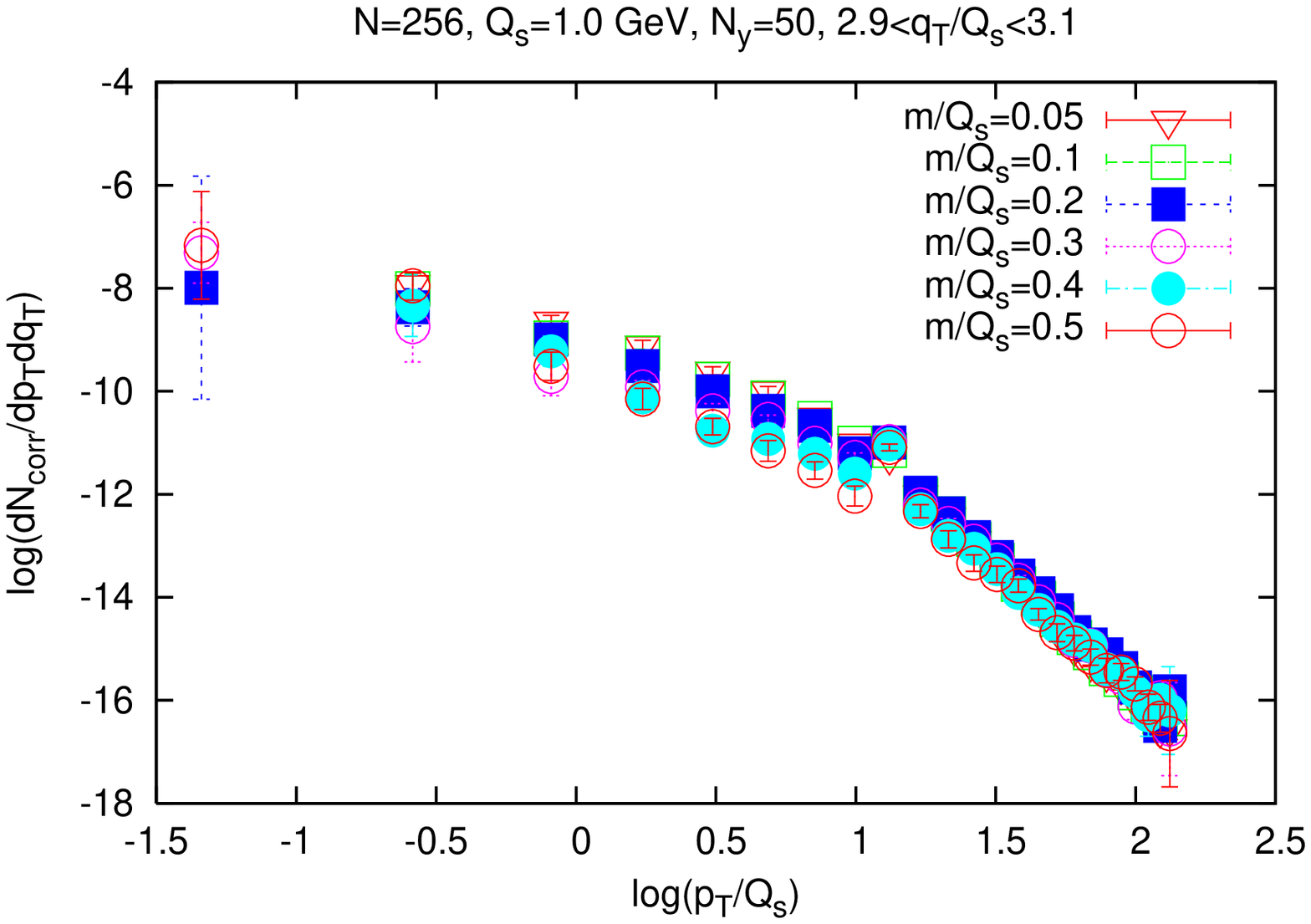}
\caption{
Sensitivity of the correlated part of the double inclusive spectrum to the value 
of IR cutoff $m/\qs$. Left: Linear plot scaled by $(p_T/\qs)^{4}$ to examine the anticipated perturbative behavior for large $p_T$. 
Right: Log-log plot of double correlated part of the double inclusive spectrum.
The results are plotted for a small bin in $q_T$ around $3\,\qs$.
}
\label{fig:double_inc_mdep_loglog}\label{fig:double_inc_mdep}
}

\FIGURE{
\includegraphics[angle=-90,width=0.45\textwidth]{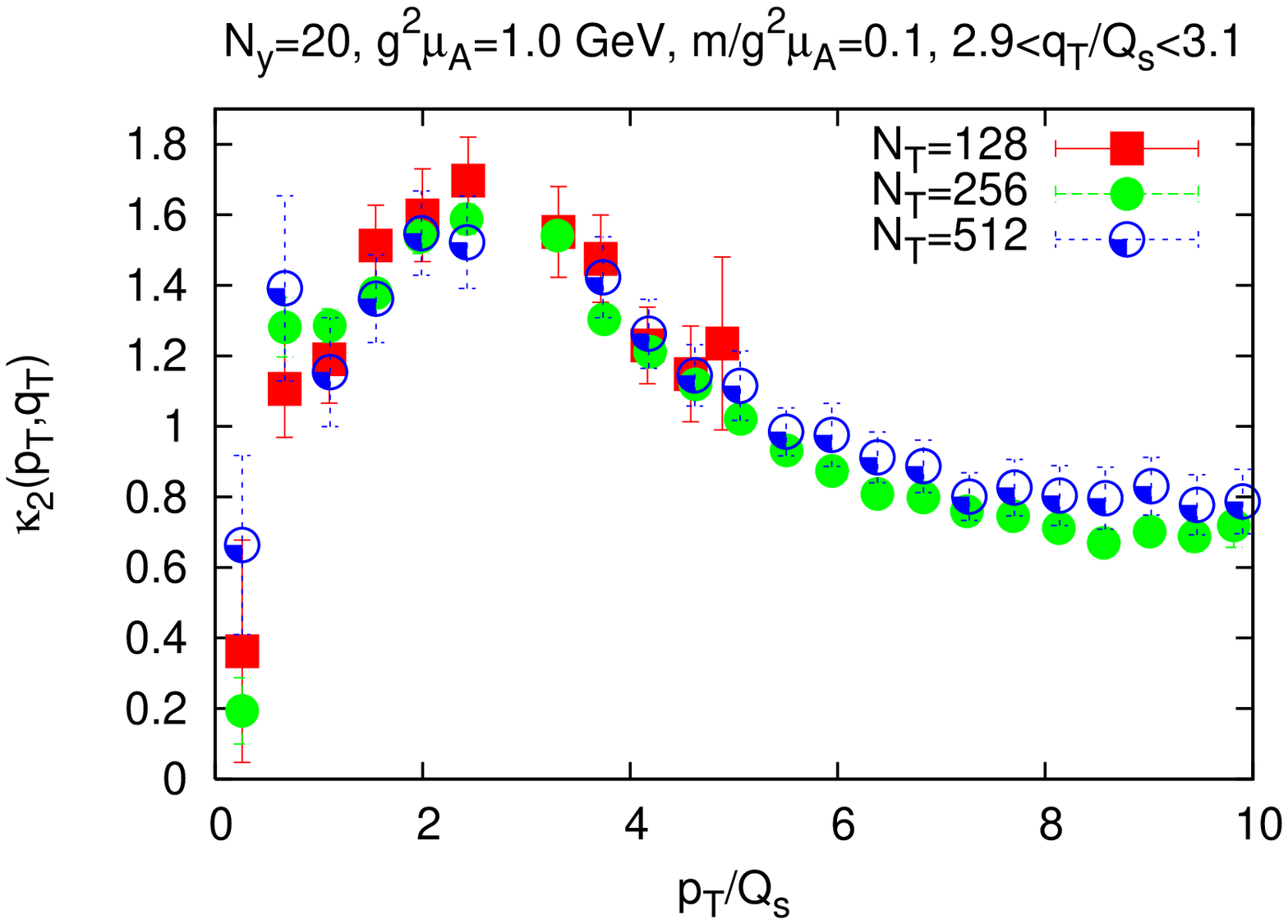}
\includegraphics[angle=-90,width=0.45\textwidth]{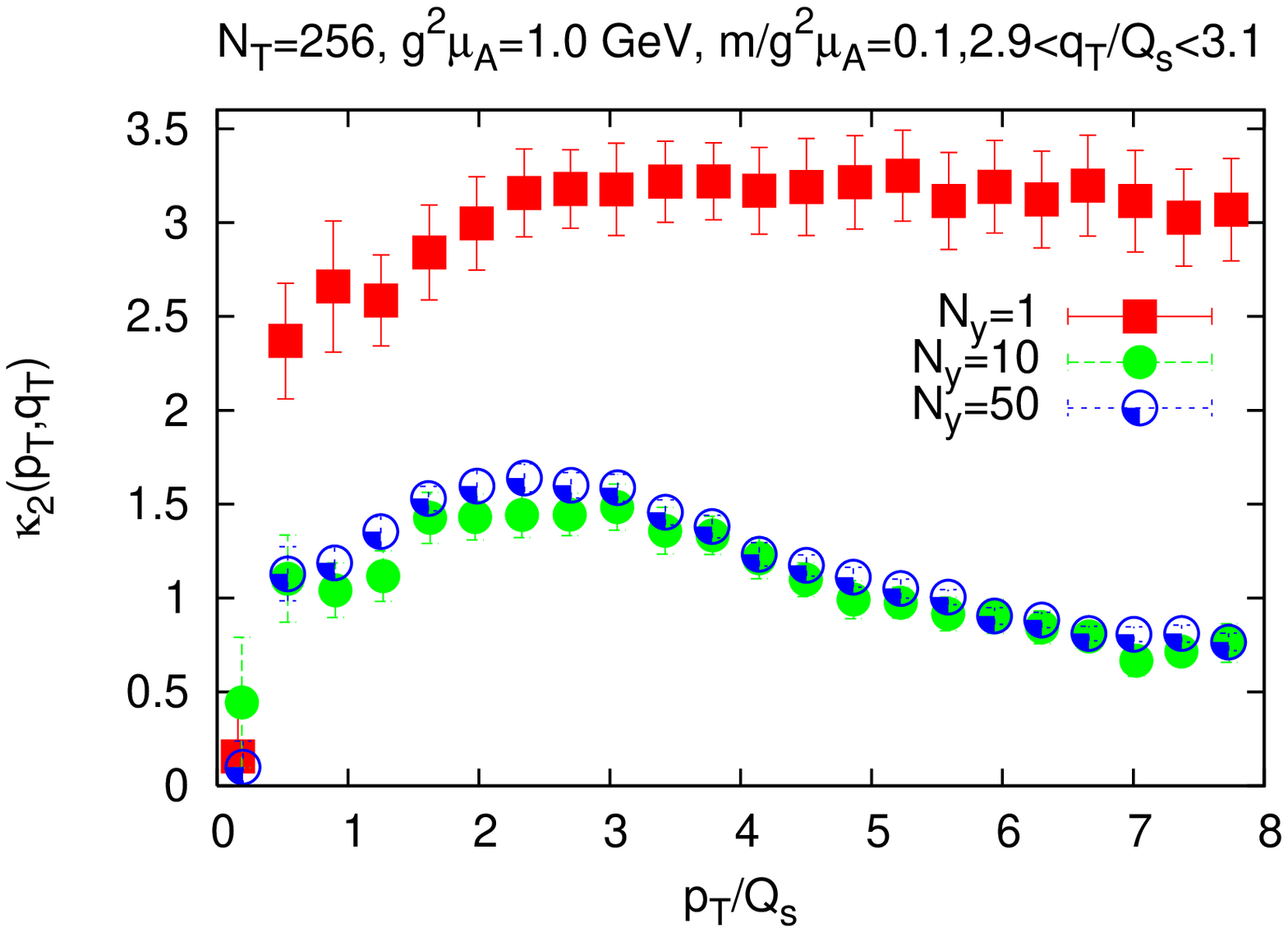}
\caption{
Left: Dependence of ${\kappa}_2$ on the transverse lattice size $N_T$ as a 
function of $p_T/\qs$ for fixed $q_T/\qs$.
Right: Dependence of $\kappa_2$ on the number of points used in the construction
of the path ordered exponential, $N_y$,
 as a function of $p_T/\qs$ for fixed $q_T/\qs$.
}
\label{fig:ntdep-double-incl}\label{fig:nydep-double-incl}
}

We now turn to the main focus of this study-the non-perturbative double inclusive 
gluon spectrum in A+A collisions.
As in the single inclusive case studied in the previous section, we will first 
examine the sensitivity of the spectrum to the lattice parameters 
spelled out in section \ref{sec:param}. Because the sensitivity of the single 
inclusive spectrum to these quantities is known to be weak, we will directly 
plot the dependence of $\kappa_2$ (see \eq\nr{eq:C2part2}) on some of these
 quantities. After exploring the sensitivity of our results to lattice 
artifacts, we will quote results for the physical value of $\kappa_2$, and 
discuss the implications of our results.

In \fig\ref{fig:double_inc_mdep}, we plot the correlated part of the double inclusive 
spectrum as a function of $p_T/\qs$, the momentum of one of the tagged gluons,
 while holding the transverse momentum $q_T/\qs$ of the other gluon fixed. 
The spectrum shown, plotted for different values of $m/g^2\mu$, is scaled by 
a factor $(p_T/\qs)^4$ in \fig\ref{fig:double_inc_mdep} (left). From the perturbative computation discussed in 
section~\ref{sec:pert}, we expect this scaled spectrum to go to a constant
 value at large $p_T/\qs$. We note that it does so approximately, keeping 
in mind that the spectrum at large $p_T/\qs$ is especially sensitive to 
lattice artifacts. We also note that the spectrum is weakly dependent on 
$m/g^2\mu$. In \fig\ref{fig:double_inc_mdep_loglog} (right), 
we also  plot the correlated double inclusive gluon spectrum as a 
log-log plot. In addition to the power law tail, we observe a qualitative change to a softer 
$p_T$ spectrum for $p_T \leq 3 \qs$.

\subsection{Determining $\kappa_2$}

In \fig\ref{fig:ntdep-double-incl}, we study 
the dependence of $\kappa_2$ on the number of transverse $N_T\times N_T$ lattice sites 
and on $N_y$, the number of points in the longitudinal direction used to
 construct the path ordered Wilson lines. In the former case, hardly any dependence is 
seen thereby indicating a rapid convergence to the continuum 
limit. For longitudinal lattices, one observes a strong dependence
 for small $N_y$, but a rapid convergence thereafter, with virtually no $N_y$ dependence 
seen for $N_y\geq 10$ onwards. The dependence of $\kappa_2$ on $g^2\mu L$ 
for a fixed value of $m/g^2\mu$ is shown in \fig\ref{fig:g2muLdep-kappa2}. As anticipated,
 the dependence of $\kappa_2$ on $g^2\mu L$ is rather weak.
This further confirms, as suggested previously, that the physics in the infrared is, for a finite $m$, 
insensitive to the size of the system $\ra$.

Now that we have established the convergence of $\kappa_2$ as a function of $N_y$ and $N_T$ is robust, 
we should remind the reader that the continuum value of $\kappa_2$, being dimensionless, is most generally 
expressed as  $\kappa_2(p_T/\qs, q_T/\qs, \Delta \varphi, \qs \ra, m/\qs)$.
In general $\kappa_2$ is  a function of two 2-dimensional vectors. When one takes into 
account rotational symmetry, it is a function of three variables and there
are many ways to plot such a function. The general structure is 
illustrated in \fig\ref{fig:3d}, where we show 
$\kappa_2$ as a function of $|\pt-\qt|$ and $|\pt+\qt|$. There is a 
delta-function  peak at $\pt = \qt$ which is clearly visible. The peak is
left out in the right hand plot of \fig\ref{fig:3d} to better illustrate 
the remaining structure. As mentioned in \se\ref{sec:numerics}, there is also an enhancement 
in the correlation in the back-to-back configuration $\pt = -\qt$, 
which can be clearly seen in  \fig\ref{fig:3d}.

\FIGURE{
\includegraphics[angle=-90,width=0.45\textwidth]{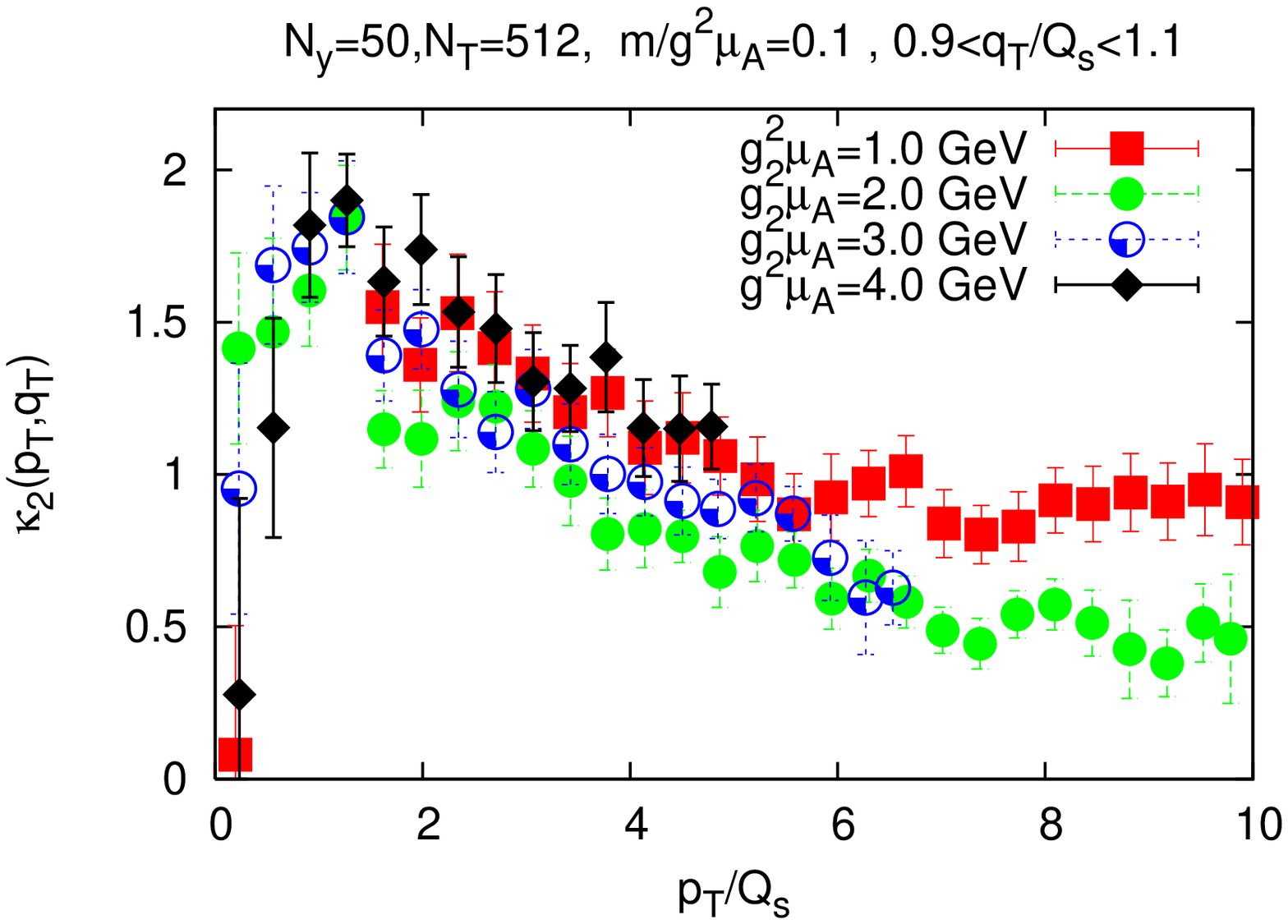}
\caption{
${\kappa}_2$ for differing values of $g^2{\mu}L$ for a fixed $m/g^2\mu$ which is equivalent here to 
a fixed $m/\qs$.}
\label{fig:g2muLdep-kappa2}
}

We now address the dependence of $\kappa_2$ on $m/\qs$.
We saw  in \fig\ref{fig:double_inc_mdep} that the double inclusive
 spectrum had a weak dependence on $m/g^2\mu$, but the
single inclusive spectrum had a slightly stronger dependence. The resulting effect on 
$\kappa_2$ is shown  in \fig\ref{fig:m_fixedqs} as a function of
$p_T$ with $q_T$ averaged over the interval $[2.9\,\qs, 3.1\,\qs]$ and 
as a function of the angle between $\pt$ and  $\qt$.  One observes that $\kappa_2$ decreases rapidly with increasing $m/\qs$ but converges 
to $\kappa_2\sim 0.5$ for larger $m/\qs$. The interpretation of the $m/\qs$ dependence of the results will 
be discussed further in the next section. 

We can establish from the r.h.s. plot in \fig\ref{fig:m_fixedqs}
that $\kappa_2$ is nearly constant as a function of 
$\Delta \varphi$; the strength of the correlation is only weakly 
dependent on the  relative azimuthal angle between pairs of gluons. This result confirms the conjecture in  Ref.~\cite{Dumitru:2008wn}.  
Turning to the dependence of $\kappa_2$ on $p_T/\qs$, $q_T/\qs$, the dependence is not 
entirely flat as assumed in Ref.~\cite{Dumitru:2008wn};
  nevertheless, despite some structure, the variation of 
$\kappa_2$ with $p_T/\qs$ is on  the order of $20$\% for $p_T/\qs \leq 4$.

\FIGURE{
\includegraphics[width=0.45\textwidth]{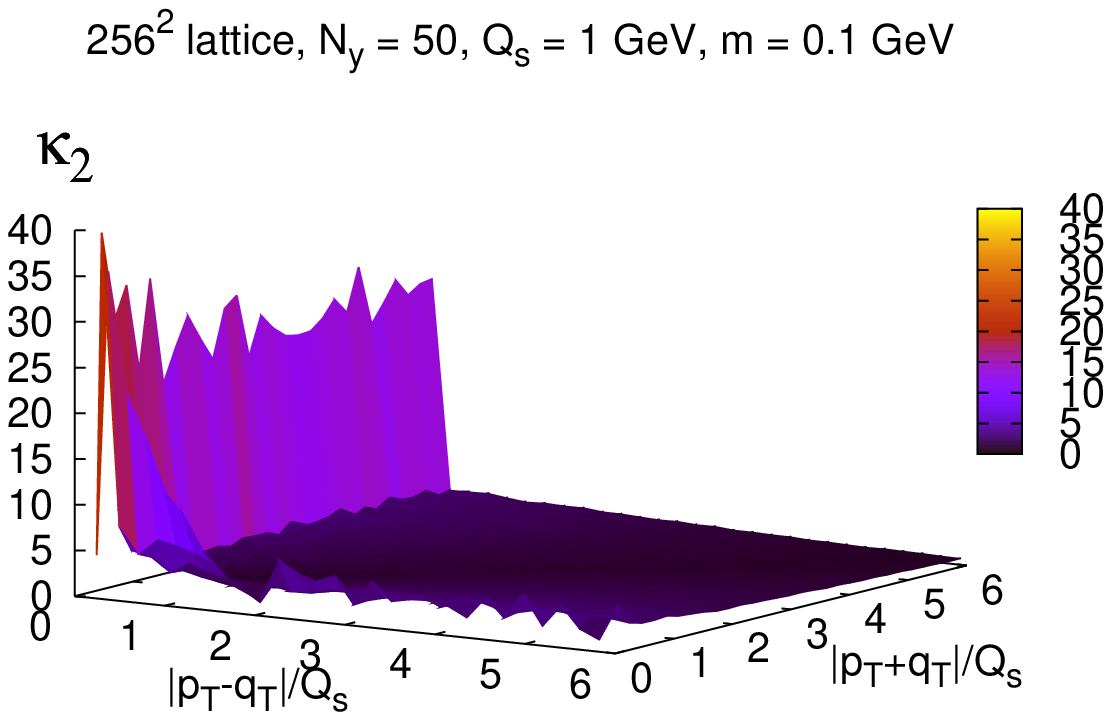}
\includegraphics[width=0.45\textwidth]{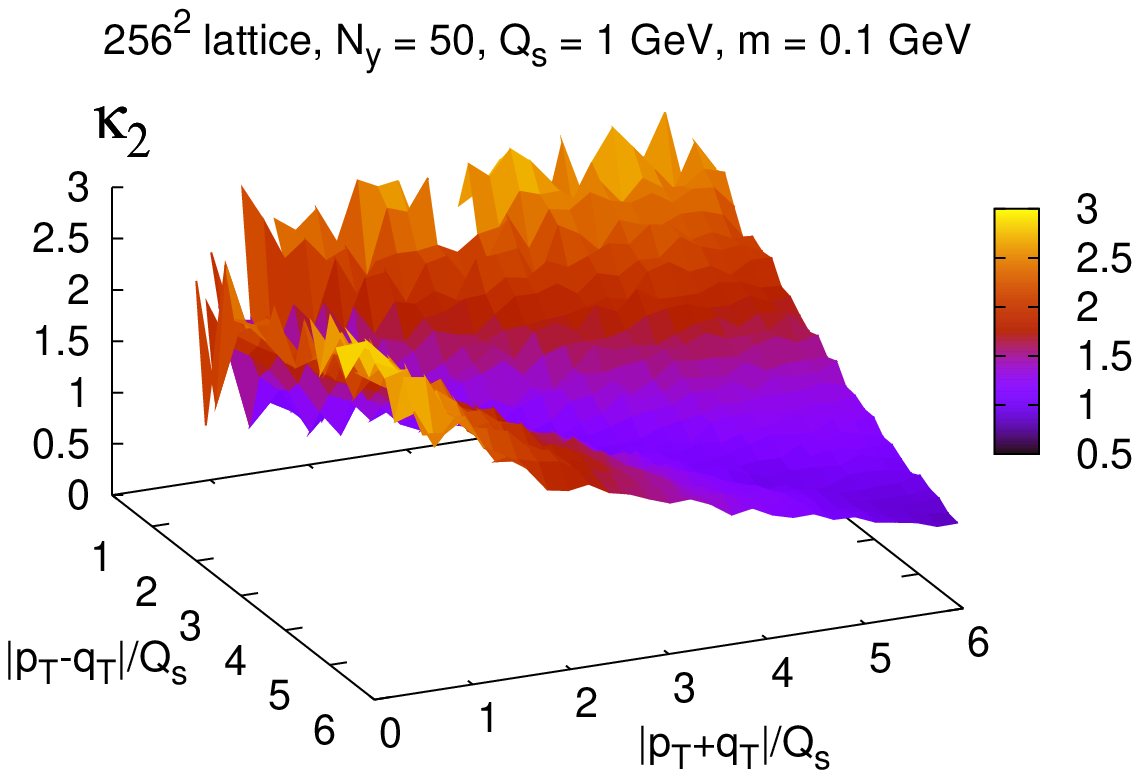}
\caption{
Three dimensional plot of ${\kappa}(|\pt-\qt|,|\pt+\qt|)$. The plot on the
left shows the delta function peak at $\pt = \qt$. On the plot on the right
this peak has been left out by reducing the range (the $|\pt-\qt|/\qs$--axis
starts from 0.1) to show the structure
around the away-side, i.e. close to $\pt = -\qt$.
}
\label{fig:3d}
}

Figures \ref{fig:qtdepvspt} and \ref{fig:pdepdiag} show 
the effect of taking different configurations of the ``trigger'' and 
``associate'' transverse momenta $p_T$ and $q_T$. In \fig\ref{fig:qtdepvspt}
(left) 
the momentum $q_T$ is fixed to three different narrow bins and $\kappa_2$ is
plotted as a function of $p_T$. Figure~\ref{fig:qtptdep} (right)
shows the dependence of $\kappa_2$ on the angle between $\pt$ 
and $\qt$ for  three different combinations
of $p_T$ and $q_T$; either both around $\qs$, both around $3\,\qs$ or one at $\qs$
and the other one at $3\,\qs$. In \fig\ref{fig:pdepdiag},
we show $\kappa_2$ for $p_T$ and $q_T$ having the same value, but 
with a finite angle between the vectors $\pt$ and $\qt$; showing a 
strong increase in the correlation for smaller values of the momenta.

\FIGURE{
\includegraphics[angle=-90,width=0.45\textwidth]{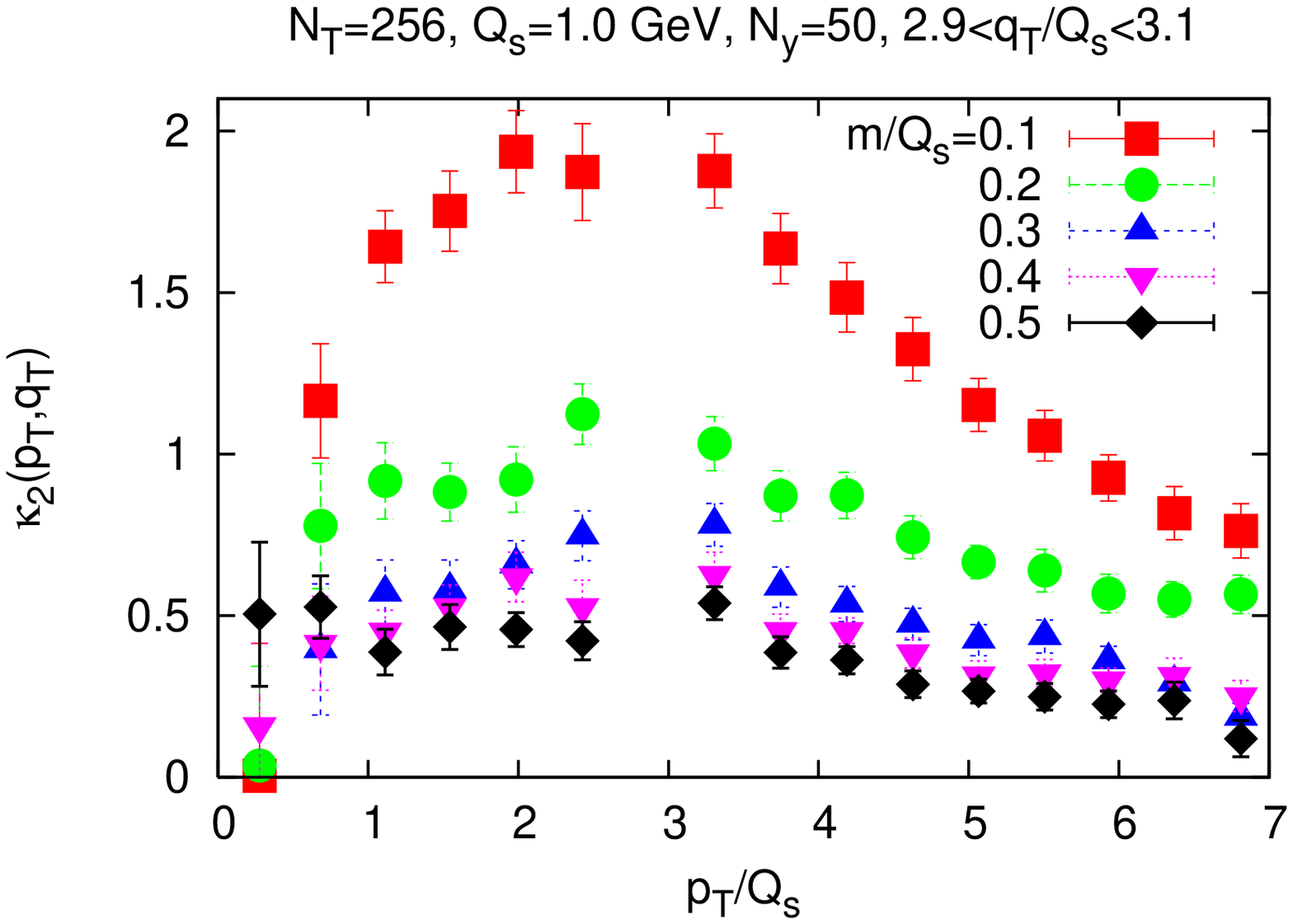}
\includegraphics[angle=-90,width=0.45\textwidth]{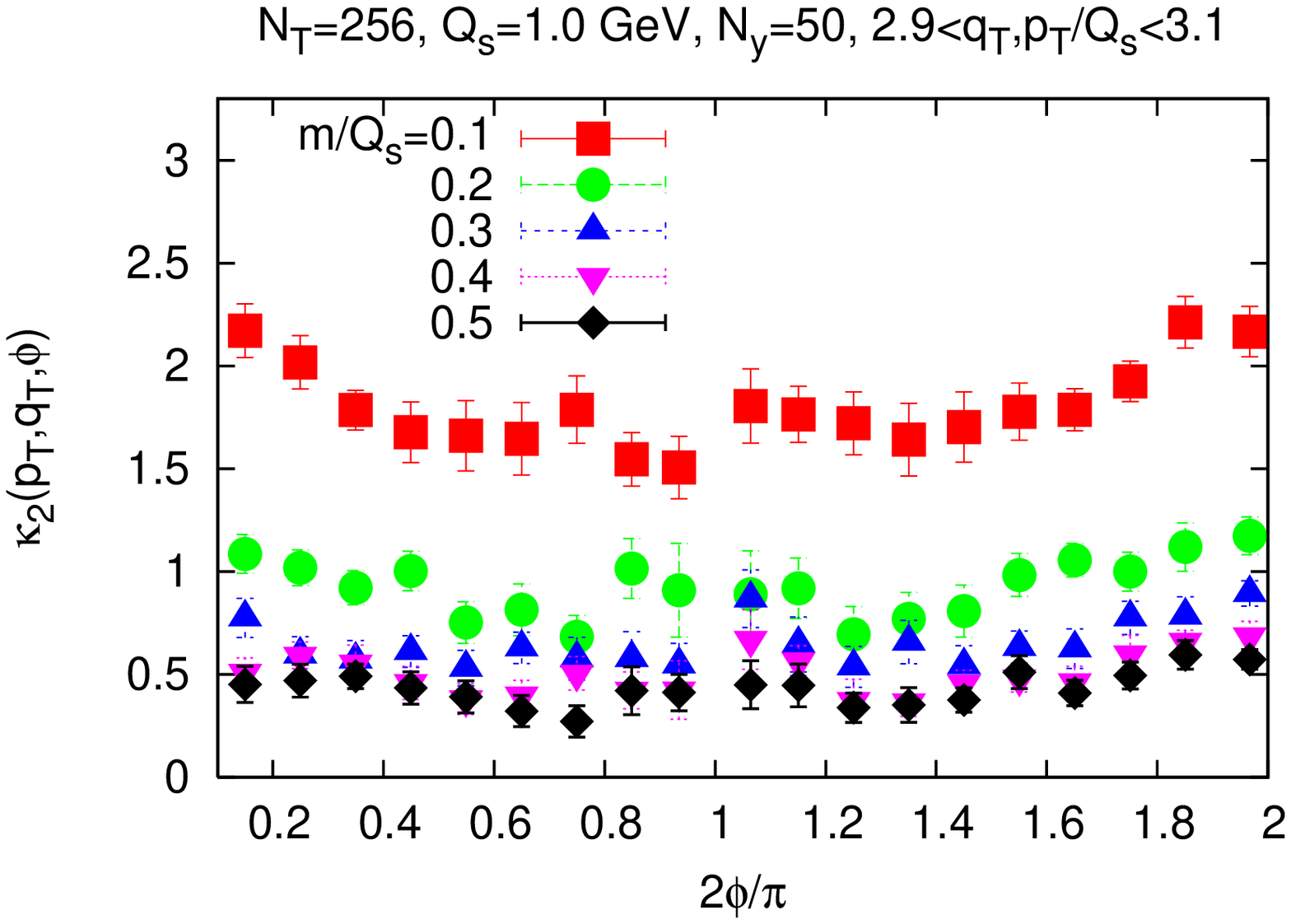}
\caption{
$\kappa_2$ for different values of $m/\qs$ with $\qs = 1 \gev$, at $N_y = 50$.
Left: 
$\kappa_2$ as a function of $p_T$ for $q_T$ in a small bin around $3\,\qs$.
Right:
$\kappa_2$ as a function of $\Delta \varphi$  with both 
 $p_T$ and $q_T$ in a small bin around $3\,\qs$.
}
\label{fig:m_fixedqs}
}

\FIGURE{
\includegraphics[angle=-90,width=0.45\textwidth]{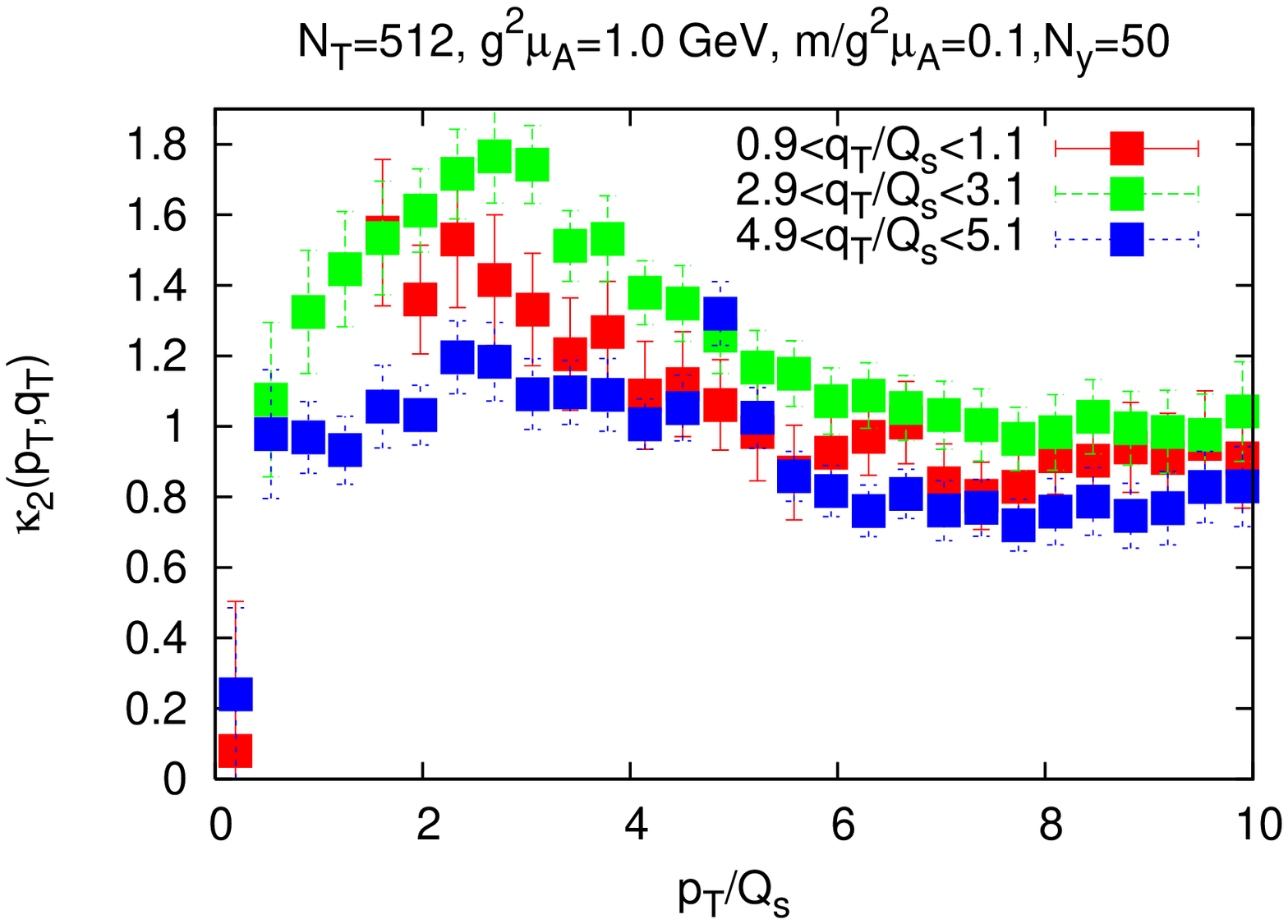}
\includegraphics[angle=-90,width=0.45\textwidth]{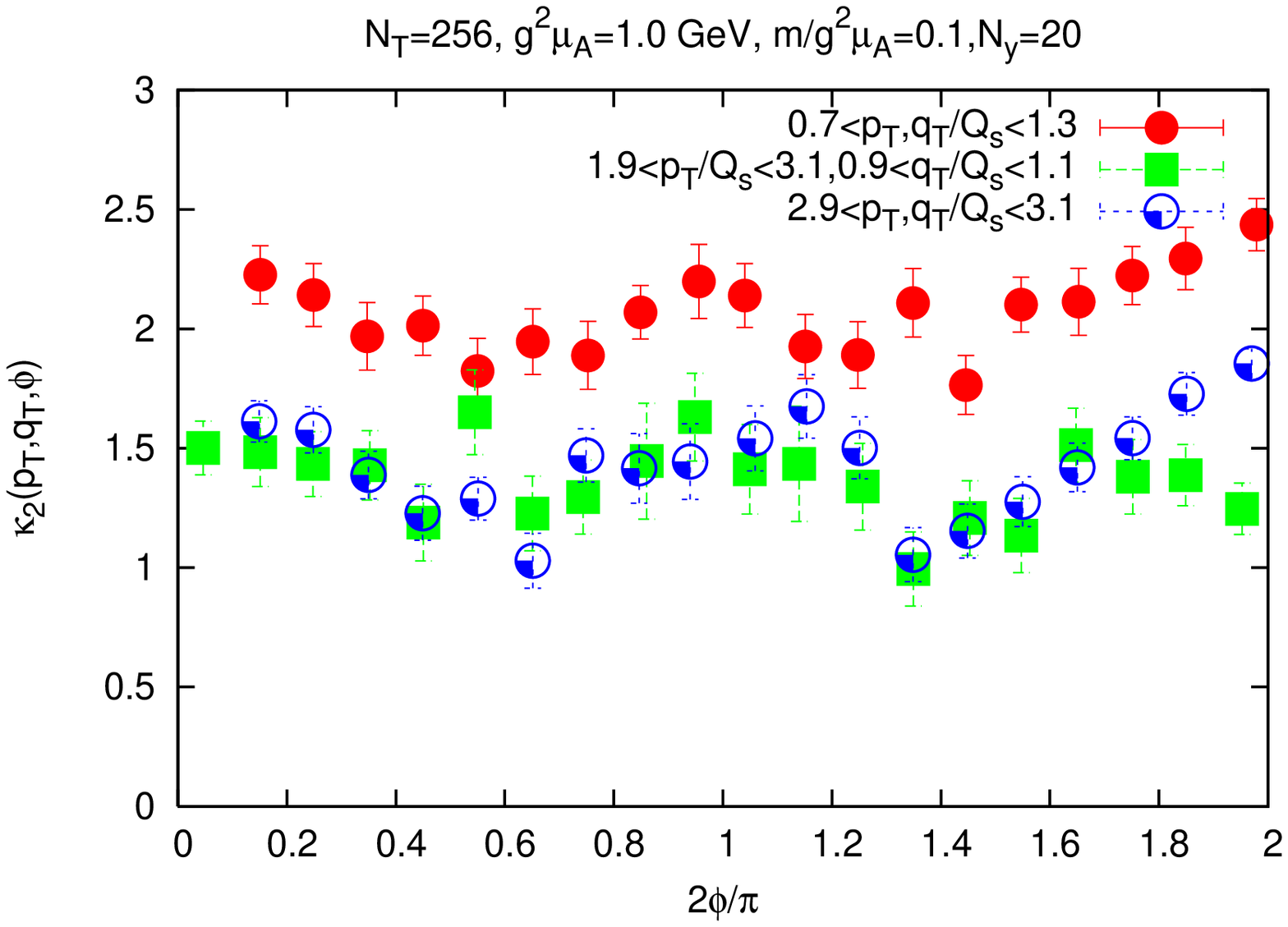}
\caption{
Left:
${\kappa}$ as a function of absolute value of one of the momenta, 
the other being fixed to the specified value.
Right:
Angular dependence of ${\kappa}$ at different absolute values of the momenta. 
Note absence of ${\delta}$-function (point at $\Delta \varphi = 0$ outside 
plotted range on the $y$-axis) at unequal momenta.
}
\label{fig:qtdepvspt}\label{fig:qtptdep}
}

\FIGURE{
\includegraphics[angle=-90,width=0.45\textwidth]{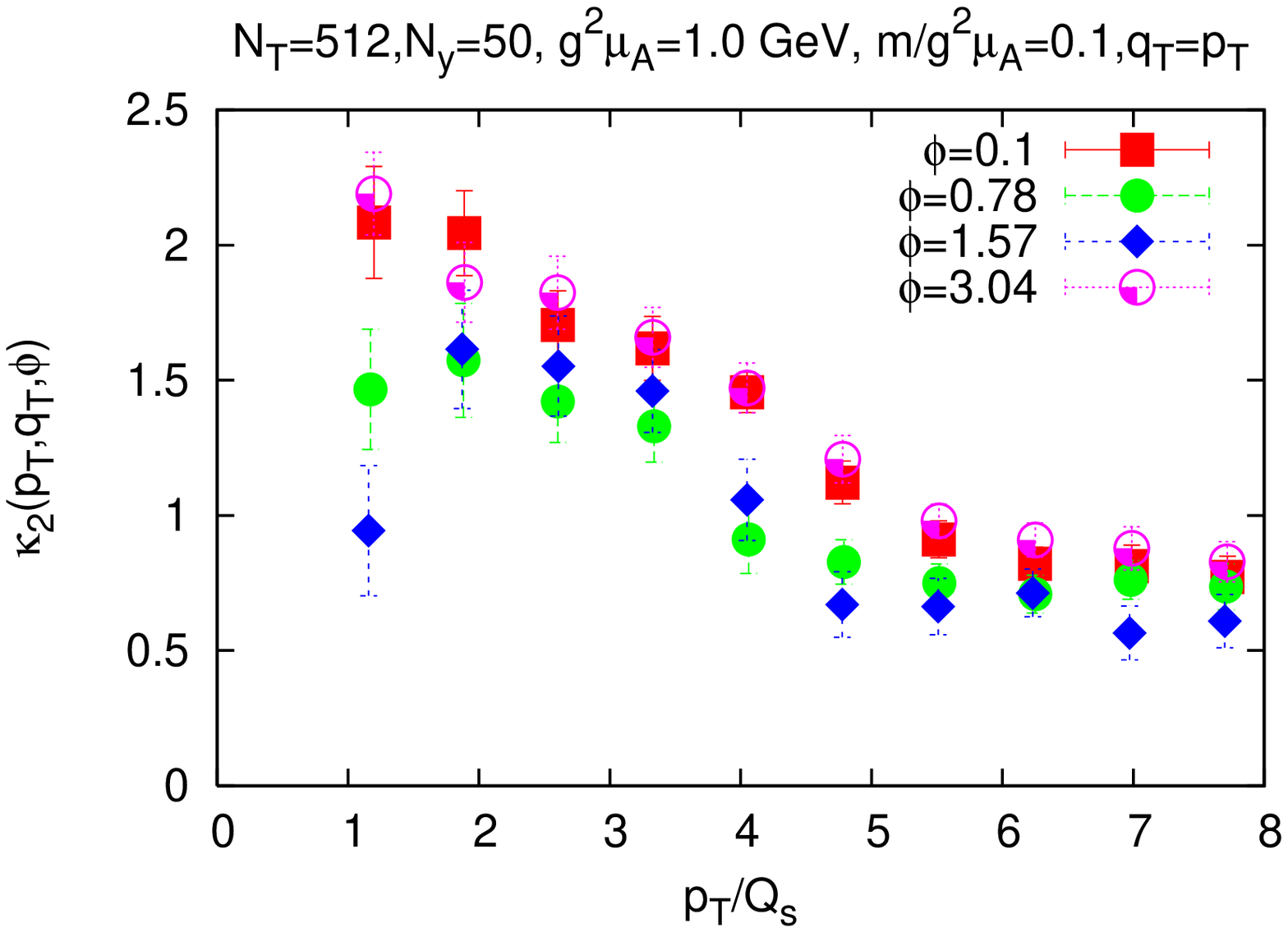}
\caption{
${\kappa}$ as function of momentum at equal momenta and specified angular separation. }
\label{fig:pdepdiag}
}

\section{Discussion and physical interpretation of results}
\label{sec:implications}

We shall now discuss the physical implications of our numerical results for the double inclusive gluon distribution in the Glasma. In the Glasma flux tube picture~\cite{Dumitru:2008wn}, as noted in \eq\nr{eq:C2part2} previously, it was conjectured that the correlated two gluon spectrum $C(\p,\q)$ can be expressed as 
\begin{equation}
\frac{C_2(\p,\q)}{
\left<\frac{\ud N}{\ud y_p \ud^2 \pt}\right>
\left<\frac{\ud N}{\ud y_q \ud^2 \qt} \right>}
=\kappa_2\,\frac{1}{S_\perp \qs^2}\,,
\label{eq:C2-discussion}
\end{equation}
where $\kappa_2$ is a non-perturbative constant which behaves parametrically as $N_c^{-2}$. The perturbative computation of $\kappa_2$ has a logarithmic 
dependence on $p_T/\qs$ and $q_T/\qs$. Modulo this logarithm, one estimates 
perturbatively\footnote{On account of incorrect factors of 2 and $\pi$, the number 
quoted in Ref.~\cite{Dumitru:2008wn} is an order of magnitude larger. The 
correct perturbative value (modulo logs) is quoted in Refs.~\cite{Gelis:2009wh,Dusling:2009ar}.} that $\kappa_2\sim 0.4$. 

In comparing our non-perturbative results to the perturbative estimate, we first 
note that $\kappa_2$ is weakly dependent on $\Delta \varphi$ and $p_T/\qs$, 
$q_T/\qs$, just as conjectured in Ref.~\cite{Dumitru:2008wn}. Our result for 
$\kappa_2$ is larger than the perturbative estimate, but it has a sensitivity, 
as shown in \fig\ref{fig:m_fixedqs}, to $m/\qs$. 
As noted previously, the double inclusive spectrum by itself 
has a relatively weak dependence on $m/\qs$ and the dependence on
$m/\qs$ in the ratio $\kappa_2$ mostly comes from the single inclusive
spectrum. Let us now discuss the physical interpretation of this
dependence.

Recall that in our computation $m$ appeared as a cutoff
inserted into \eq\nr{eq:uprod} to invert the Laplace operator. It regulates
the long distance Coulomb tails of the color field\footnote{The propagator for the 2~dimensional Laplace operator is a logarithm.}.
In the context of the MV model, this scale appears as an additional external 
parameter, whose value is not determined within the model itself.
Thus it is not completely clear whether the 
scale $m$ should be thought of as the confinement scale or a scale 
of order $\qs$. In a straightforward implementation it is natural to 
think of $m$ as a cutoff imposing color neutrality at the size of a nucleon. This is indeed the picture used for instance in Refs.~\cite{Krasnitz:2002ng}. 

This picture does not, however, take 
into account color screening effects coming from quantum evolution 
effects (virtual and radiative corrections) that are described by 
the JIMWLK and BK equations. At RHIC energies they may not be very
 important but will be extremely important at LHC energies. The 
effect of color screening of the correlators has been considered
 previously~\cite{Iancu:2002aq,*Mueller:2002pi} (see also the 
discussion in  Ref.~\cite{Gelis:2009wh}) and it is argued that 
quantum evolution effects  regulate the infrared behavior of the 
color charge density correlator
at a scale that is also parametrically of 
the order of $\qs$. This is not a
 purely classical effect arising from non-linear Yang-Mills 
dynamics but from a combination of rescattering and quantum
 evolution effects. To determine the ``correct'' value of $m$ 
in this context  it is thus necessary to go beyond
the MV model and include the effects of high energy evolution. 
A systematic numerical study of the two gluon correlation when 
high energy evolution effects are included is beyond the scope 
of this work.

A practical way to address this question in our simulation is to note
 that for a geometrical flux tube picture to be valid the two particle 
cumulant should be given by the size of the flux tube
$S_\textrm{ft}$ (representing 
the transverse range of color correlations) divided by the transverse 
area of the system times a number of order unity.
We can make this logic explicit by expressing the size of the 
typical flux tube as 
\begin{eqnarray}
\frac{C_2(\p,\q)}{
\left<\frac{\ud N}{\ud y_p \ud^2 \pt}\right>
\left<\frac{\ud N}{\ud y_q \ud^2 \qt} \right>}
=\frac{S_\textrm{ft}}{S_\perp}\,, 
\end{eqnarray}
where $S_\textrm{ft} = \kappa_2 (\cdots,m/\qs)/\qs^2$. For the 
range of $m/\qs = 0.1$--$0.5$ considered, $\kappa_2 \approx 2$ - $0.5$ converging 
rapidly to the latter value for increasing $m/\qs$. The effective 
scale governing correlation strengths of unit strength is then 
$1/\sqrt{S_\textrm{ft}}\approx 0.7$--$1.4\,\qs$. This scale,
although not numerically very large, is nevertheless  
a semi-hard scale and corresponds to transverse distances
 much smaller than the nucleon size, thereby confirming the picture 
of early times in the Glasma as classical field configurations
that are coherent over transverse distances much smaller than a nucleon. 

Let us then turn to a comparison with the RHIC data on 
two particle correlations. The experimentally observed quantity is 
$\Delta \rho/\sqrt{\rho_{\rm ref.}}(\Delta \varphi)$; in our 
framework, for $\Delta \varphi=0$, it can be expressed as  
\begin{eqnarray}
{\Delta \rho\over \sqrt{\rho_{\rm ref.}}} \left(\Delta \varphi =0\right) &=& {dN\over dy} \cdot \frac{C_2(\p,\q)}{
\left<\frac{\ud N}{\ud y_p \ud^2 \pt}\right>
\left<\frac{\ud N}{\ud y_q \ud^2 \qt} \right>}\, (\gamma_B - \frac{1}{\gamma_B})\nonumber \\
&=& {K_N\over \as}\, (\gamma_B - \frac{1}{\gamma_B}) ,
\label{eq:flux-data1}
\end{eqnarray}
where $K_N = \kappa_2 /13.5$ for an SU(3) gauge theory and $\gamma_B$ is the average radial boost in the framework of a blast wave model. From the 
RHIC data~\cite{Adams:2004pa,Daugherity:2008su}, one can estimate that $\Delta \rho/\sqrt{\rho_{\rm ref.}}(\Delta \varphi=0)= 1/\sqrt{2\pi \sigma_\varphi^2}$, with 
$\sigma_\varphi=0.64$. Combining this with \eq\nr{eq:flux-data1}, one obtains 
\begin{equation}
\kappa_2^{\textrm{BW}} \sim {0.7 \over (\gamma_B - \frac{1}{\gamma_B})} \, ,
\end{equation}
for $\as =0.5$ and where the superscript denotes that this is a crude 
estimate extracted from experiment using a blast-wave parametrization. 
For an average blast wave radial velocity $V_r =0.6$, this gives 
$\kappa_2^{\textrm{BW}}\sim 1.5$; 
for $V_r =0.7$, one obtains $\kappa_2^{\textrm{BW}} \sim 1$. There is 
considerable variation therefore in the value of $\kappa_2^{\textrm{BW}}$ obtained 
from the ridge data due to the final state flow parametrization. 

An alternative method to extract $\kappa_2$ is to compare the expression 
for the \emph{negative binomial} multiplicity distribution 
(cf. \eqs\nr{eq:n-cumulant}--\nr{eq:n-cum-k} in the 
Glasma~\cite{Gelis:2009wh} to PHENIX data~\cite{Adare:2008ns} on 
the same. From this comparison, one obtains
\begin{equation}
\kappa_2^\textrm{NBD} \sim 3.9.
\label{eq:flux-data2}
\end{equation}
One sees therefore considerable variation in the extraction of $\kappa_2$ from experiment within the present framework. Within the many uncertainties, 
one can say at best it is a number of order unity. Our study suggests that a coherent picture of such correlations from RHIC and higher LHC energies can 
in principle provide unique information on color screening of strong fields at early times in heavy ion collisions.

An objective of this study is also to proceed in the opposite direction, namely, to determine $\kappa_2$ from a non-perturbative computation and use this as input into a dynamical space-time evolution model. As guide to this future program is the work in Ref.~\cite{Takahashi:2009na}, where two particle correlations are extracted from the hydrodynamical evolution of flux tube structures in A+A collisions. 

\section{Summary}

In this paper, we investigated the validity of the Glasma flux tube
 scenario of multi-particle correlations by performing a non-perturbative
 numerical computation of double inclusive gluon production in the Glasma.
 Our results were obtained by solving Yang-Mills equations on a space-time
 lattice for Gaussian distributed color source configurations with a 
variance proportional to the saturation scale $\qs^2$. Our results confirm 
key features of the Glasma flux tube scenario. Particles produced from 
coherent longitudinal electric and magnetic fields in transverse regions 
of size $\sim 1.4/\qs$--$1/2\,\qs$,  much smaller than the 
nucleon size, have correlations of unit strength. As in the asymptotic 
perturbative estimates, particles produced from the flux tubes are 
uncorrelated in the relative azimuthal angle between the particles,
the observed azimuthal collimation being produced later from radial 
flow.  The correlations show non-trivial structure in $p_T$ and $q_T$, 
which smoothly go over to perturbative results in the limit of
 $p_T/\qs\gg 1$, $q_T/\qs \gg 1$. Our results for the two particle
 correlation strength are consistent with estimates of the strength
 extracted from model comparisons to RHIC data. 
A useful extension of this work is to incorporate our results into
 more detailed hydrodynamical models of the space-time evolution of
 initial state correlations. 

\section{Acknowledgments}
We thank Adrian Dumitru, Sean Gavin, Larry McLerran
and Jun Takahashi for very helpful discussions. We are especially 
grateful to Fran{\c c}ois Gelis for numerous useful remarks and 
insights that have significantly improved this work. S.S.'s research
 is supported by a SCIDAC grant and by a Lab Directed Research 
\& Development (LDRD) grant from Brookhaven National Laboratory. 
R.V.'s research is supported by DOE Contract No.  DE-AC02-98CH10886. 
T.L. is supported by the Academy of Finland, project 126604. 

\appendix
\section{Lattice formulation}

The Yang--Mills equations can be formulated as Hamilton's equations of motion. To preserve gauge invariance, they are solved numerically 
on a lattice where the degrees of freedom at a site $i$ are the link variables
\begin{equation}
    U_{i}(\xt) = \exp\left[ i g a A_i(\xt) \right],
\end{equation}
where $a$ is the lattice spacing on a transverse lattice. The appropriate discretized lattice (Kogut-Susskind) Hamiltonian in our case is given by 
\begin{equation}\label{eq:kogutsusskind}
aH = \sum_{\xt} \Bigg\{ \frac{g^2 a}{\tau}\trace E^iE^i +
\frac{2\nc\tau}{g^2 a} \left( 1-\frac{1}{\nc}\R \ \trace U_\Box \right)
+\frac{\tau}{a} \trace \pi^2 +
\frac{a}{\tau} \sum_i
\trace \left(  \phi - \tilde{\phi}_i \right)^2 \Bigg\},
\end{equation}
where $E_i,U_i,\pi\textrm{ and }\phi$ are 
dimensionless lattice fields that are matrices in color space, with
$E^i = {E_a}^i t_a$ etc. and the generators of the fundamental representation
normalised  as $\trace (t_a t_b) = \delta_{ab}/2$.

The first two terms are the transverse electric and magnetic fields with
the transverse plaquette defined as 
\begin{equation}
U_\Box(\xt) = U_x(\xt)U_y(\xt+\ex)U^\dag_x(\xt+\ey)U^\dag_y(\xt).
\end{equation}
The last two terms are the kinetic energy and covariant derivative of the rapidity component of
the gauge field $\phi \equiv A_\eta = -\tau^2 A^\eta$. The latter 
becomes an adjoint representation scalar following the assumption of boost invariance.
For the parallel transported scalar field, we use the notation 
\begin{equation}
\tilde{\phi}_i(\xt) \equiv U_i(\xt)\phi(\xt+ \ii)U_i^\dag (\xt).
\end{equation}
In the Hamiltonian, \eq\nr{eq:kogutsusskind}, there is a residual invariance under 
gauge transformations depending only on the transverse coordinates. 

The Hamiltonian  equations of  motion are 
\begin{eqnarray}\label{eq:mo1}
\dot{U}_i &=& i \frac{g^2}{\tau}E^i U_i \nosum{i},
\\
\label{eq:mo2}
\dot{\phi} &=& \tau \pi,
\\
\label{eq:mo3}
\dot{E}^x &=& \frac{i \tau}{2 g^2} \left[U_{x,y}+U_{x,-y} - \hc \right]
- \textrm{trace} 
+ \frac{i}{\tau} [\tilde{\phi}_x,\phi] 
\\
\nonumber
\dot{E}^y &=& \frac{i \tau}{2 g^2} \left[U_{y,x}+U_{y,-x} - \hc \right]
- \textrm{trace} 
 + \frac{i}{\tau} [\tilde{\phi}_y,\phi] ,
\\
\label{eq:mo4}
\dot{\pi} &=& \frac{1}{\tau}\sum_i\left[ 
\tilde{\phi}_i + \tilde{\phi}_{-i} - 2\phi \right].
\end{eqnarray}

Gauss's law, conserved by the equations of motion, is given by 
\begin{equation}\label{eq:gauss5}
\sum_i
\left[U^\dag _i(\xt-\ii)E^i(\xt-\ii)U_i(\xt-\ii)  
-  E^i(\xt)\right]  \nonumber \\
- i [\phi,\pi] = 0\,.
\end{equation}

On the lattice the initial conditions \nr{eq:initcond} are 
\begin{eqnarray}\label{eq:latinitcond}
0 &=& \trace \left[t_a \left(\left(U^{1}_i +U^{2}_i\right)
\left(1+U_i^\dag\right) - \hc \right) \right], \label{eq:init1} \\ 
E^i &=& 0, \label{eq:init2} \\
\phi &=& 0, \label{eq:init3} \\
\pi(\xt) &=&  \frac{-i}{4g} \sum_{i} \bigg[
\left(U_i(\xt) - 1\right)
\left(U_i^{\dag 2}(\xt)-U_i^{\dag 1}(\xt) \right)
\\ \nonumber & +& 
\left(U_i^{\dag}(\xt-\ii) - 1\right)
\left(U_i^{2}(\xt-\ii)-U_i^{1}(\xt-\ii) \right) -\hc
\bigg], \label{eq:init4}
\end{eqnarray}
where $U_i^{1,2}$ in \eq\nr{eq:init1} are the link matrices corresponding to 
the color fields of the two nuclei ($A_i^{1,2}$ in \eq\nr{eq:ainit}). 
The link matrix $U_i$, which corresponding to the $\tau \ge 0 $ 
color field $A_i$, is determined by solving \eq\nr{eq:init1}.

\bibliographystyle{JHEP-2modM}

\bibliography{spires}

\end{document}